# Combined certainty and uncertainty across development frees phenotypic variation in evolution

Yue Zhang[1]*

[1] Institute of Biotechnology, University of Helsinki; Helsinki, 00014, Finland
* Email: renaissance.of.truth.chasing@gmail.com

**Abstract:** Developmental bias plays a major role in phenotypic evolution. Some researchers have argued that phenotypes, regulated by development, can only evolve along restricted trajectory under certain scenarios, such as the case for mammalian molar size ratios. However, this view has been challenged. Broadly speaking, sources for phenotypic variation remain largely unknown. The study here presents a generalized Inhibitory Cascade Model and explains that the original model described only means of phenotypes resulting from selection when viewed under a higher taxonomic scope. Consequently, I propose the combined property of development: certainty, when the prior intersegmental inhibition is strong, and uncertainty, when the opposite holds. This property potentially not only explains counterintuitively high levels of developmental instability, but also plays an essential role in generating phenotypic variation.

.

**Main Text:**

At a higher level, the field of evolutionary developmental biology (evo-devo) is intrinsically connected with organicism, a classical branch of metaphysics led by philosophers over thousands of years, including Plato (*1*), Kant (*2*), and many others. In contrast to its counterpart, mechanism, in organicism objects are perceived as a whole instead of an aggregation of deductible parts (*3*). Under such a thinking framework, scientists need to look beyond genes for a better understanding of evolution (*4*). Although marginalized during the heyday of Modern Synthesis (*5*), the field of evo-devo has made substantial progress over the past few decades, both theoretically and empirically (*6-8*). Breakthrough discoveries have been emerging from studies on model organisms/tissues (*8-11*). Among those, mammalian teeth are particularly interesting because of the easy accessibility in both natural and lab settings and the significance in evolution. Placental mammals typically have three molars: M1, M2, and M3. The Inhibitory Cascade Model (ICM) found that the phenotypic variation pattern across murine species is consistent with that of the mutant mouse teeth *in vitro* regarding molar size ratios, the size of M2 being the arithmetic mean of M1 and M3 (*10*). This finding has subsequently been viewed as an example showing development regulating evolution to move along a specific trajectory (*12*). Nevertheless, empirical studies across various mammalian groups have suggested that the ICM equation may have limited applicability beyond murines, with stronger deviation more commonly found under the intraspecific scale (*13-20*). The interplay between development and selection on generating phenotypic variation among molar sizes, or among morphological features in general, remains unknown (*12*). More than 164 years after "*The Origin of Species*", "*Our ignorance of the laws of variation is*", still, "*profound*" (*21*) (p. 167).

The study here first presents a generalized ICM, and then discusses the combined property of certainty and uncertainty across developmental stages, and its vital role in generating phenotypic variation. Specifically, Section I presents a generalized ICM based on developmental data. Section II presents the covariation patterns among intermolar inhibitions as revealed by anthropoid teeth of natural samples. Section III presents broader applications of the generalized ICM, along with the mathematical formula. Section IV presents intermolar covariation patterns of teeth with heterogeneous shapes/functions, and how the various patterns are connected with selection. Section V presents the concepts and properties of dev-space and dev-type. After that, Section VI summarizes the key property of development: a combination of certainty and uncertainty across earlier to later stages, and the final section discusses how this hypothesis explains the discrepancy between lab mutants and natural forms on achieving morphological complexity.

## I. Re-examination of the ICM based on the original experiment

### I-1. The generalized ICM

Suppose that during the molar developmental process, at time $T_i$, the sizes of the three molars are $M1_i, M2_i,$ and $M3_i$. At a later time $T_j$, the molar sizes become $M1_j, M2_j,$ and $M3_j$. The size differences, using the later to deduct the earlier, are denoted as $\delta M1, \delta M2,$ and $\delta M3$, respectively. Given a molar $M_s$, with $s \in \{1, 2, ...\}$, the inhibition from $M_s$ to $M_{s+1}$ is denoted as $I_s$. Then, to quantify the intensity of inhibition, $I_s = \frac{\delta M_s}{\delta M_{s+1}}$.

According to the experiment leading to the ICM, intermolar inhibition gets accumulated from anterior to posterior (*10*), meaning $I_2 > I_1$. Hence, $1/I_2 < 1/I_1$, which equals to

$$\frac{\delta M_3}{\delta M_2} < \frac{\delta M_2}{\delta M_1}$$

(1)

The original ICM equation, M2 = ½ (M1+M3), was derived from the means of molar sizes, which only captured the mean signal from the final state of the developmental experiment (*10*). In contrast, Inequality (1) here is the quantitative expression of ICM's principle, that is, intermolar inhibition gets accumulated from anterior to posterior.

### *I-2. Relationship and covariation between $I_1$ and $I_2$*

According to the rationale presented above, slopes from the linear regressions of the sizes of M3 on M2, and M2 on M1 can be used to quantify $1/I_2$ and $1/I_1$, respectively. The *in vitro* developmental data (see materials and methods and tables S1 and S2) (*10*) reveal that during the explant developmental interval, $1/I_2$ is almost consistently lower than $1/I_1$, agreeing with Inequality (1) (Fig. 1A). Furthermore, comparing samples between different lab treatments (*10*), for the explants with post-M1 lamina detached, earlier versus later, from M1 (i.e., E13 cut vs E14 cut), both $I_1$ and $I_2$ decreased. On average, $I_2$ decreased by a larger degree than $I_1$ (Fig. 1B), consistent with the original findings reported by Kavanagh and colleagues (*10*). How $I_2$ covaries with $I_1$ is further analyzed using evolutionary data in the following sections.

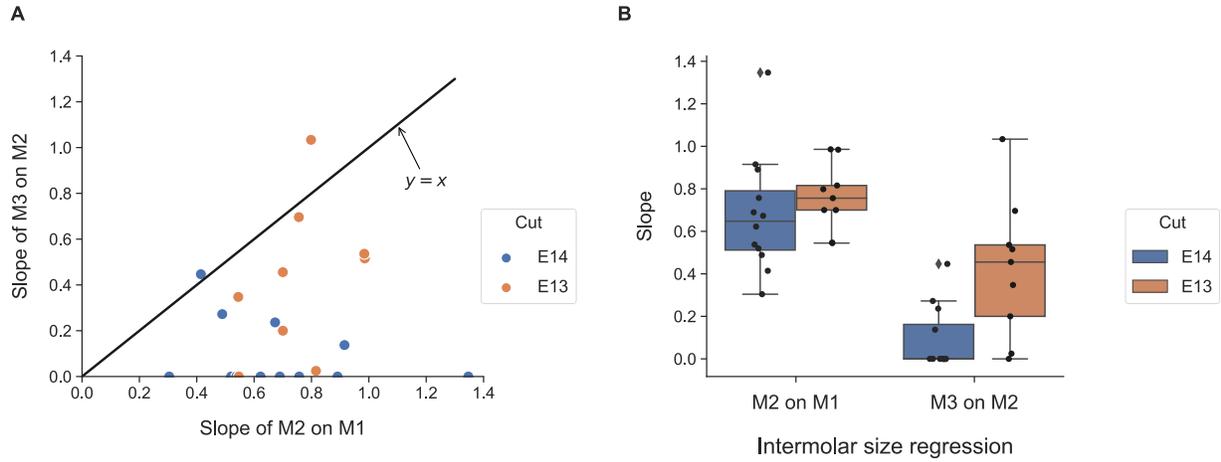

**Fig. 1. Developmental data agreeing tightly with the generalized ICM**. (**A**) the scatterplot of the slopes of linear regressions of M2 on M1 and that of M3 on M2 (see tables S1 and S2 and figs. S1 and S2). (**B**) boxplots of the slopes.

## II. Covariation between intermolar inhibitions revealed by natural samples

### II-1. Rationale of using intermolar size slope to infer intermolar inhibition

The developmental data from specimens *in vitro* show that the intensity of intermolar inhibition stays relatively stable for each individual throughout the cultured growth period, as supported by the high goodness-of-fit (as measured by the value of $r^2$) of the linear regressions (figs. S1 and S2, table S2). In contrast, for evolutionary data, directly measuring intermolar inhibition is not practical. Nevertheless, there is still a way to infer inhibition intensities and related parameters. To interpret this rationale using mathematical expressions, here suppose that given *n* individuals, the developmental initiation time and ending time are denoted as *T0* and *T1*, respectively, with the instant growth rate at time $t$ ($t \in [T0, T1]$) denoted as $r(t)$, noting that the growth rate is a function of time *t*. Then, regarding a specific tooth, such as M1, for a given individual *g*, $g \in \{1, 2, \ldots, n\}$, its developmental initiation time and ending time are denoted as $T0_{g,M1}$ and $T1_{g,M1}$, respectively, and the instant growth rate at time $t$ is $r(t)_{g,M1}$. Hence, the size of M1 for this individual can be calculated as $\int_{T0_{g,M1}}^{T1_{g,M1}} r(t)_{g,M1} \, dt$.

Accordingly, given a population, if there exists such a relationship $\frac{\delta M_2}{\delta M_1} = C$ ($C$ is a constant), then, using the notation defined above, for any two individuals within the population, denoted as *i* and *j*, respectively, there is:

$$\frac{\int_{T0_{j,M2}}^{T1_{j,M2}} r(t)_{j,M2} \, dt - \int_{T0_{i,M2}}^{T1_{i,M2}} r(t)_{i,M2} \, dt}{\int_{T0_{j,M1}}^{T1_{j,M1}} r(t)_{j,M1} \, dt - \int_{T0_{i,M1}}^{T1_{i,M1}} r(t)_{i,M1} \, dt} = C$$

(2)

Equation (2) indicates that when any two individuals are compared, as M1 increases (or decreases), M2 increases (or decreases) proportionally. This linear relationship suggests that the M1-to-M2 inhibition intensities across individuals within this population are not expected to deviate far from *1/C*. It is important to note that Equation (2) does not imply any relationship among growth rates, instant or averaged, or developmental durations of the individuals in comparison. Below, anthropoid molar data from Plavcan's dissertation (*18, 22*) are analyzed (see materials and methods).

### II-2. The scale-dependent issue of linear regressions

First, the intermolar size regressions are analyzed on the species level. Unlike the case for developmental data, the value of $r^2$ varies substantially among species. As the taxonomic unit under analysis goes up (e.g., species to tribe, subfamily, or family), the value of $r^2$ turns significantly higher (fig. S3). Such a clade-dependent pattern on linear regressions is commonly observed in studies of ecology (*23, 24*).

Below, the size regressions of M3 on M1 are examined to further explain this issue. As expected, the linear fit of M3 on M1 is strong if and only if both of that for M3 on M2 and for M2 on M1 are strong (fig. S4A). Additionally, the slope of M3 on M1 measures the compound effect of $I_1$ and $I_2$, because the slope approximates $1/I_1 I_2$ (fig. S4B). A corresponding ecological index, the size ratio of M3/M1, is also examined here to interpret the biological implications of the scale-dependent pattern. As Lucas and colleagues have pointed out, the size ratio of M3/M1 is informative for revealing the dietary composition of anthropoids (*25*).

Across above-species clades analyzed here, Colobinae, Cercopithecini, and Papionini, within each clade on the species level, there is a weak to moderate negative correlation between the standard deviation of M3/M1 and the $r^2$ values from the size regression of M3 on M1 (fig. S5). Statistically, it suggests that as the spread of M3/M1, or the span of ecological niches, goes broader, the slope for M3 on M1 becomes more divergent in general, and vice versa. When the clade is viewed as a whole, however, both values are significantly higher compared to the situation within individual species (fig. S5). This may seem counterintuitive but can be explained with further visualization on the KDE (Kernal Density Estimation) plots of the whole clade compared with that of separate species (figs. S6-S8). In contrast to the relatively smooth and flattened pattern shown at the clade level, within each individual species, the distribution of M3/M1 shows stronger signals of discontinuity and multimodality, but these gaps are filled in when other related species are included. In other words, the niche span is more disrupted at intraspecific level versus more continuous at higher taxonomic scales. This explains the linear fit difference between intra- and inter- specific levels (fig. S3). Hence, although the linear fit is higher as scale goes up, the slope only reflects the aggregated signal such as in the case here (figs. S9-S11). Accordingly, whenever possible, analysis should be conducted at the finest scale (*24*).

### II-3. The utility trade-off among M1, M2, and M3 of various anthropoid taxa

Because of the inhibitory cascade, from the developmental perspective, an increase or decrease of $I_1$ leads $I_2$ to change towards the same direction. It can be related to the functional trade-off of M1 versus M3, with M2 acting as the medium. From the functional perspective, as Lucas and colleagues (*25*) have explained, a wider jaw with larger P4/M1 and smaller posterior molars is more adapted for processing relatively big and sticky food particles, whereas a longer jaw equipped with molars of subequal size is more adapted for processing relatively small and non-sticky ones, and larger tooth areas are adapted for processing abrasive food.

Before discussing how intermolar inhibitions covary across various taxa due to the functional trade-off, the expected utility, originated from Daniel Bernoulli's hypothesis on the St. Petersburgh's Paradox (*26, 27*) is briefly introduced here. In general, using $w$ and $U(w)$ to denote wealth and its utility, then we can expect $U'(w) > 0$ and $U''(w) < 0$. $U'(w) > 0$ indicated that the utility is positive, and $U''(w) < 0$ indicates that the gain of marginal utility diminishes as utility increases.

Using α and β to denote the slope of M2 on M1 and that of M3 on M2, respectively, covariation patterns between intermolar inhibitions across anthropoid taxa are analyzed below. First, most of the slopes fall beneath the line $y = x$, indicating that the inhibition from M2 to M3 is consistently stronger than that from M1 to M2, agreeing with the generalized ICM (Fig. 2).

Second, β can be viewed as a utility function of α, denoted as $U(α)$. When only the mean values are examined, it is seen that M3 is absent (β = 0) when α ≈ 1/2. Going from there, as α increases, β increases at a faster pace, until they reach the point α ≈ β ≈ 1 (Fig. 2). This trajectory, which connects Callitrichini, Cebidae, and Colobinae, satisfies the equation $U(α) = 2 - 1/α$, corresponding to the ICM line M2 = ½ (M1+M3) (supplementary text). For the increasing speed of β relative to that of α, $U'(α) = 1/α^2$, noting that $1/α^2 > 1$ when α ∈ [1/2, 1). Hence, functionally speaking, as the inhibition from M1 to M2 decreases, the marginal gain of M3 is higher than the marginal loss of M1 (using M2 as the gauge), and vice versa. This property is consistent with the finding that diet generalists such as Cerbidae are located in the middle, whereas diet specialists such as Callitrichini and Colobinae are found at the lower and upper boundary, respectively. In addition, clades do not follow this curve when placed outside of the interval (Fig. 2B). Noting that when α > 1, there is $U'(α) < 1$, thus the marginal gain of M3 is lower than the marginal loss of M1 outside of the interval. Accordingly, it is expected that only highly specialized herbivores, such as ungulates, are located along or even above the curve when α > 1. Finally, this utility curve, with the specifically constrained interval, has also been found within murines (*10*). Anthropoids and murines, both belonging to the superorder Euarchontoglires, share the same property that molars are homogeneous in shape.

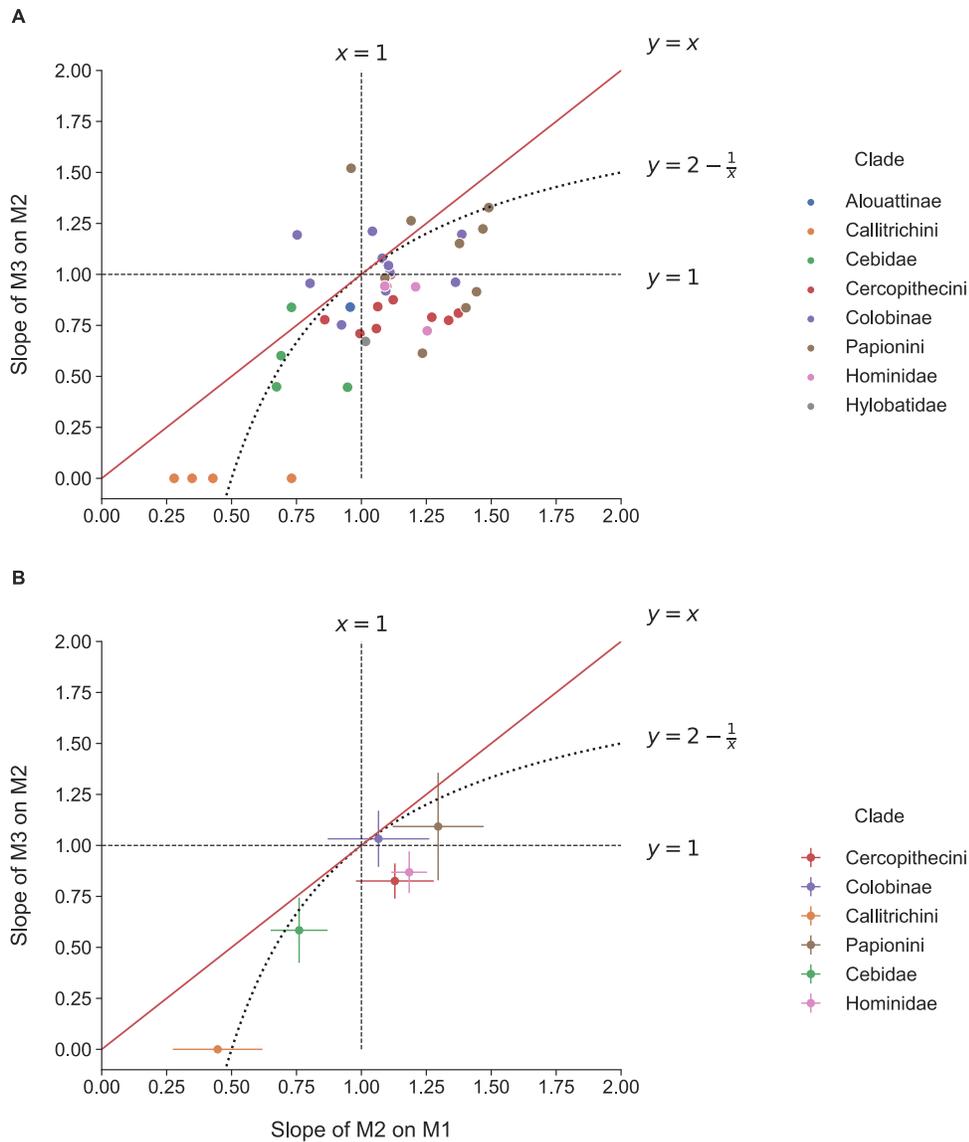

**Fig 2. Covariation between the slopes resulting from regressions of adjacent molars for anthropoids.** Data are from Plavcan *(20, 22)* (table S3). The red line indicates the upper boundary of the slopes resulting from regressions of M3 on M2, as suggested by the generalized ICM. The dashed curve indicates the utility curve as discussed in the text. (**A**) Slopes of each species, differentiated by clade. (**B**) The mean and standard deviations of each clade (clades containing more than one species in the dataset).

### III. The generalized math expression for covariation among intersegmental inhibitions

*III-1. Broader applications of the generalized ICM*

Below, applicability of the generalized ICM is explored. As shown in fig. S12, ratios of mean molar sizes could potentially be used to roughly infer intersegmental inhibitions if the slope data were not available for the case of anthropoids here (supplementary text). Out of convenience, size ratios reported by previous studies are used in the tests below. It is important to note that, however, mean size ratios, fundamentally different from slopes, are only used as a secondary option when data for the primary option (slopes) are not available due to insufficient sampling. For stylopod, zeugopod, and autopod of hindlimbs, using the data across 1452 population-averaged measurements of 1354 species complied by Young (*28*) (see materials and methods), it is seen that the covariation pattern overwhelmingly follows the generalized ICM (fig. S13), agreeing with the finding of Young that the proportion of zeugopod was more frequently found to be larger than 1/3 (*28*).

Beyond animals, the generalized ICM principle may also be applicable to plants. In a recent study on bipinnately compound leaves of *Serianthes nelsonii*, Deloso and Marler (*29*) tested the plasticity of leaf morphometrics across various light retreatments. Under any case, a leaflet in the middle cannot be considerably smaller than those both before and after (Fig. 2 in (*29*)). Their finding supported the generalized ICM presented here, which suggests a more common occurrence of convex or straight rather than a concave outline (see Section III-2 below). It is noted that developmental regulation on the origination and growth of leaflets, though may be better studied compared to animal segments, is far from thoroughly known (*30-34*), and thus drawing direct connection between these two categories is premature at the current stage. Nevertheless, the covariation pattern shown in leaflets of *Serianthes nelsonii* suggests that the generalized ICM may be applicable to plants to some extent, awaiting more evidence from future studies.

Despite the potentially broad applicability of the generalized ICM, the model should not be regarded as universally restrictive. For example, sizes of phalanges, of which the middle can be significantly smaller than both before and after (*35*), seem to violate the generalized ICM. More studies are needed to better understand those violations.

*III-2. Implication on convex to straight outlines instead of concave ones for organisms*

Following the generalized ICM, considering a simplified situation in which all segments are homogeneous in shape, then a convex or straight, rather than concave, outline is more likely to be formed (Fig. 3, fig. S14, and supplementary text). This is an oversimplified interpretation with imprecise substitutions applied, examples across both extant and extinct organisms, however, do suggest an overwhelming occurrence of convex and straight versus concave outlines along chained segments. The generalized ICM, along with many other factors, such as other models derived from Turing's morphogenesis model (*36*), may partially explain this phenomenon.

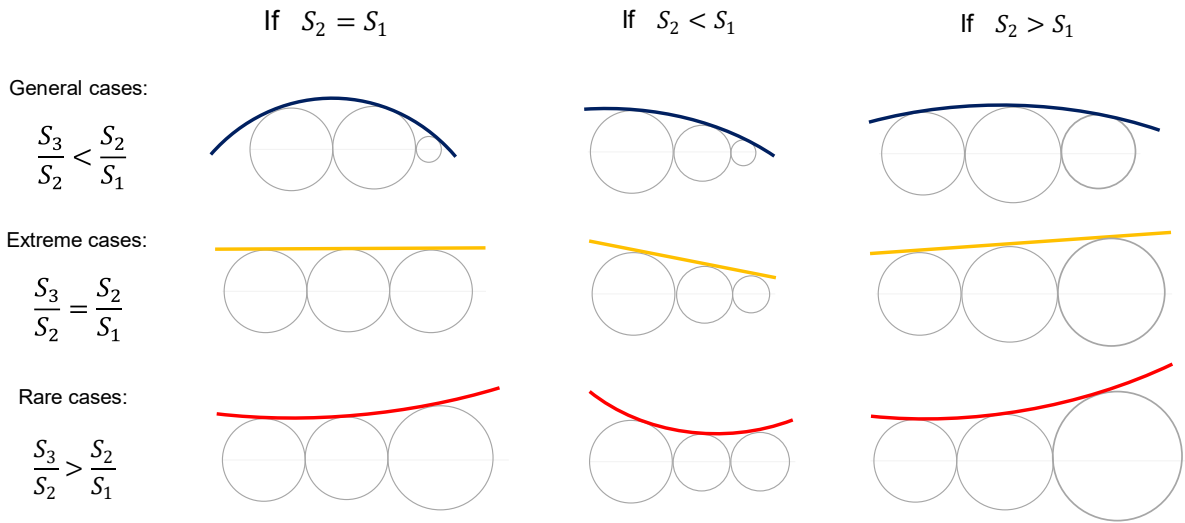

**Fig 3. Simplified sketch showing various expected outlines for homogeneous segments under corresponding conditions.**

*III-3. Mathematical expression of the generalized ICM*

As a general situation, for developmental segments/structures that share the same developmental domain (e.g., cusps on the same tooth, teeth along the same lamina), let us denote them as $S_1$, $S_2$, ..., $S_n$. Then, the generalized ICM indicates:

$$\frac{\delta S_{i+2}}{\delta S_{i+1}} < \frac{\delta S_{i+1}}{\delta S_i} \tag{3}$$

Here, $\delta S_{i+2}$, $\delta S_{i+1}$, and $\delta S_i$ denote the growth magnitude of each segment/structure, respectively, with $i \in \{1, 2, ..., n-2\}$. Then, according to the pattern revealed by both experimental data (Fig. 1) and evolutionary data (Fig. 2), we can further conclude that,

$$\frac{\delta S_{i+2}}{\delta S_{i+1}} = \begin{cases} 0, & if \ \frac{\delta S_{i+1}}{\delta S_i} \leq T_i \\ V_u, \ V_u \in [0, \frac{\delta S_{i+1}}{\delta S_i}), & if \ \frac{\delta S_{i+1}}{\delta S_i} > T_i \end{cases} \tag{4}$$

In Equation (4), $T_i$ denotes the threshold value. $V_u$ denotes an undetermined value. The equation indicates that if the inhibition intensity from the earlier developed pair is strong enough, then no additional segment/structure can be developed at the shared growth domain. Conversely, the third segment/structure may or may not initiate, and if initiating, then the relative growth is undetermined, but no higher than the capped value as regulated by the immediately preceding segment/structure pair. It is noted that the capped value, as indicated by Equation (4), is derived from tests on segments grown along a single main axis. It is expected that additional parameters (e.g., newly induced coefficients) will be required when the underlying cases become more complicated, but the general principle should hold.

### IV. Utility curves for molars with heterogeneous shapes/functions

Analysis on the anthropoid data indicates that covariation among adjacent segments of homogeneous shape/function tends to follow a specific utility curve within a restricted interval. In comparison, molars of carnivorans, which are heterogeneous in shape and function, are analyzed here. This analysis, including the data used (see materials and methods), is built upon prior studies contributed by Asahara (*19, 37*) and Asahara and colleagues (*20*), who pointed out coordinated covariation between M1's talonid/trigonid (Tal/Tri) and molar size ratios.

*IV-1. Overview on shared threshold values and various utility curves*

Four carnivoran families, which typically possess distinctive carnassial teeth, are analyzed here (Fig. 4). For most felids, the Tal/Tri ratio is below the threshold for M2's initiation, but this should not be the case for all felids, as the presence of M2 has been reported for Eurasian lynx

*(38)*. For the other three families, the Tal/Tri ratio is above the threshold as M2 is present even for species with the lowest Tal/Tri ratio (0.18). Hence, the threshold should be lower than 0.18, but above the ratios of the felids with no M2. For mustelids, despite the positive correlation between Tal/Tri and M2/M1, a property shared among carnivorans (*20*), the ratio of M2/M1 is capped at a threshold value (0.31), below which M3 cannot initiate (Fig. 4). Notably, this threshold value is shared between canids and mustelids, as the two canid species with below-threshold M2/M1 ratios, bush dog (*Speothos venaticus*) and dhole (*Cuon alpinus*), have no M3 developed. Hence, this threshold value for the initiation of M3 may be shared across carnivorans. In addition, as indicated by Equation (4), exceeding the threshold value does not guarantee the initiation of the later developed segment. For instance, although the Tal/Tri ratio of honey badger (*Mellivora capensis*) exceeds the threshold value for M2's initiation, M2 is normally absent (*39*).

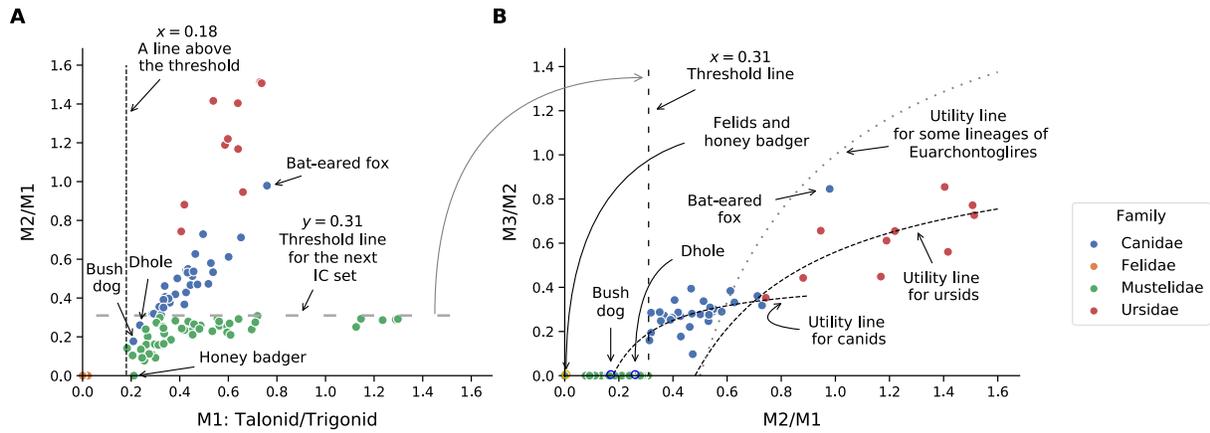

**Fig 4. Comparisons of covariation among intermolar (including molar parts) size ratios across various carnivoran taxa**. Data are from Asahara and colleagues (*20*). (**A**) Covariation between M2/M1 and talonid/trigonid of M1. (**B**) Covariation between M3/M2 and M2/M1. The utility curve of canids follows the equation $y = 0.45 - \frac{0.08}{x}$, and the utility curve of ursids follows the equation $y = 1.08 - \frac{0.52}{x}$, according to prior studies (*19, 20*).

Compared to canids, ursids are more omnivorous in general, of which grinding is more important. As reflected in the utility curve for covariance of M3/M2 on M2/M1, the curve of ursids is steeper than that of canids, indicating that using M2 as the gauge, as the relative size of M1 decreases, the marginal gain of M3 decreases at a lower rate in ursids versus canids. The utility curve, derived from the functional trade-off of slicing versus grinding (*25*), is not only clade specific but also scale dependent, as the curve of carnivorans is expected to approximate the weighted average of those from each family including canids and ursids. For carnivorans, because animal matters generally take a higher proportion in food, a longer molar row is needed less. Hence the utility curve looks significantly flattened compared to that of euarchontaglires. In contrast, euarchontaglires do not possess molar size patterns with M2 significantly smaller than M1 along with a highly reduced, or absence of, M3, which is developmentally reachable. Such a form, however, would be adaptively inferior to a carnivoran taxon with corresponding molar size ratios, and thus was selected against.

### *IV-2. Utility curve being clade specific but phylogenetically unconstrained*

Bat-eared fox (*Otocyon megalotis*) is traditionally recognized as being insectivorous, with termites being the most common food source (*40*). There is also evidence suggesting that bat-eared fox should be recognized as a generalist feeding on a variety of food resources depending on seasonality (*41*). The molar morphology of bat-eared fox, resembling an archaic form of tribosphenic tooth with no distinctive carnassial features, suggests a balanced functional trade-off between slicing and grinding, which is adapted for a generalized opportunistic feeding strategy. Notably, although belonging to canids, the molar size ratios of bat-eared fox follow the utility curve of euarchontaglires instead of canids. It shows that, albeit being clade-specific, utility curves are determined by functional adaptation rather than phylogenetic constraints.

### *IV-3. Rarity of M4 for terrestrial placental mammals primarily due to selection not development*

Bat-eared fox regularly possesses four molars in the lower jaw (*40*). Although this is an exceptional case for terrestrial placental mammals, the M3/M2 ratio (or the slope of M3 on M2 as a more accurate metric linking with development) exceeds the threshold for initiation of M4 in a variety of species. The appearance of M4 has been frequently documented since studies dating back from the late 19$^{th}$ century (*42, 43*). In some taxa, such as orangutans, the occurrence rate of M4 may reach up to 20% in the sample reported (*43*). Among carnivorans, M4 may even emerge when M3 reaches only one third of the size of M2, as in the case of some coyote specimens (*37*). Therefore, the rarity of M4 in terrestrial placental mammals is attributed to being selected against in most cases, rather than developmental barrier.

### V.    Higher variability of intermolar inhibition at the later developmental stage and its implication on evolution

Above discussed about the generalized ICM, along with various molar size covariation patterns in relation to selection. In this section, developmental variability and the connection with adaptation are analyzed in further detail.

*V-1. Comparison among Old World monkeys*

From the middle-late Miocene to Pliocene and Pleistocene, climate in Africa first went through a phase with strong aridification and then a phase with high volatility (*44, 45*). Fossil records suggest that earlier colobines were more diversified in diets and locomotion compared to their extant counterparts (*46, 47*). Ecological changes have driven the lineage to become more specialized in folivory over the course. In contrast, papionines have maintained a wider dietary spectrum and a broader habitat and geographic range throughout their evolutionary history (*46, 47*). As for molar sizes, when the three clades of Old World monkeys, Colobinae, Cercopithecini, and Papionini, are compared, relative to the disparity among M1-to-M2 inhibitions, the M2-to-M3 inhibition shows much higher variability, with a broad variation range occupied by papionines, fully covering that of the other two clades (fig. S15). This finding suggests that the posterior pair of intermolar inhibition, while being partially dependent on the anterior pair, bears greater variability and can be a determinant factor for morphological disparity.

*V-2. Brief remarks on the importance of developmental variability on human evolution*

It has been found that early humans also lived in open savannah, sharing similar diets and habitats with baboons, and facing similar challenges brought by the highly fluctuating environment (*48*). Yet, the highly variable ecological setting acted as a critical catalyst for human evolution (*45*). Accessibility to a variety of food resources, especially meat, was achieved early in the evolution of human lineage, well before the appearance of *Homo erectus* (*49, 50*). Increased consumption of meat and high-quality food has been found to be crucially linked with enlargement of brain, increased usage of tools, and improved cognition in human evolution (*50, 51*). With these advantages, later evolved species (e.g., *homo erectus, homo neandertals, homo sapiens*) along the human evolutionary lineage were able to cope with environmental volatility and food scarcity more efficiently. These species have a significantly higher level of dental fluctuating asymmetry compared to earlier hominine species and apes (*52*). Continuing with this trend, even with relatively stable food supply (e.g., post-agricultural), modern humans still maintain a high level of disparity on dental traits. This is reflected not only from within-local-population samples (*53*), but also from within-individual samples (*54, 55*). Overall, at each level, from within-individual to among-population and beyond, diversification of phenotypes is essential, because it provides more opportunities for surviving through environmental fluctuations (*56*). Below, I introduce the concepts of dev-space and dev-type, in relation to but distinct from morphospace and morphotype, and discuss the role of developmental variability in evolution.

*V-3. Variability of dev-types*

Using the case of molar size inhibition as an example, given the evolutionary data sample of a species or a higher taxon, the dev-types can be inferred from the size regressions of molars (Fig. 5A). The developmental potential, though, occupies a broader area than the expressed dev-types. Without much change on the earlier developmental parameter, all the reachable values for the later developmental parameter should be counted as potential dev-types, such as the area below the $y = x$ line for the case here. Correspondingly, given a seeding population, if new niches for more leaves/grass feeding emerge, with selection favoring higher usage of M3, then new dev-types will initiate above the original ones. Conversely, if new niches for less leave/grass feeding and more fruit/meat feeding emerge, with selection favoring higher usage of M1, then new dev-types will initiate below the original ones (Fig. 5B).

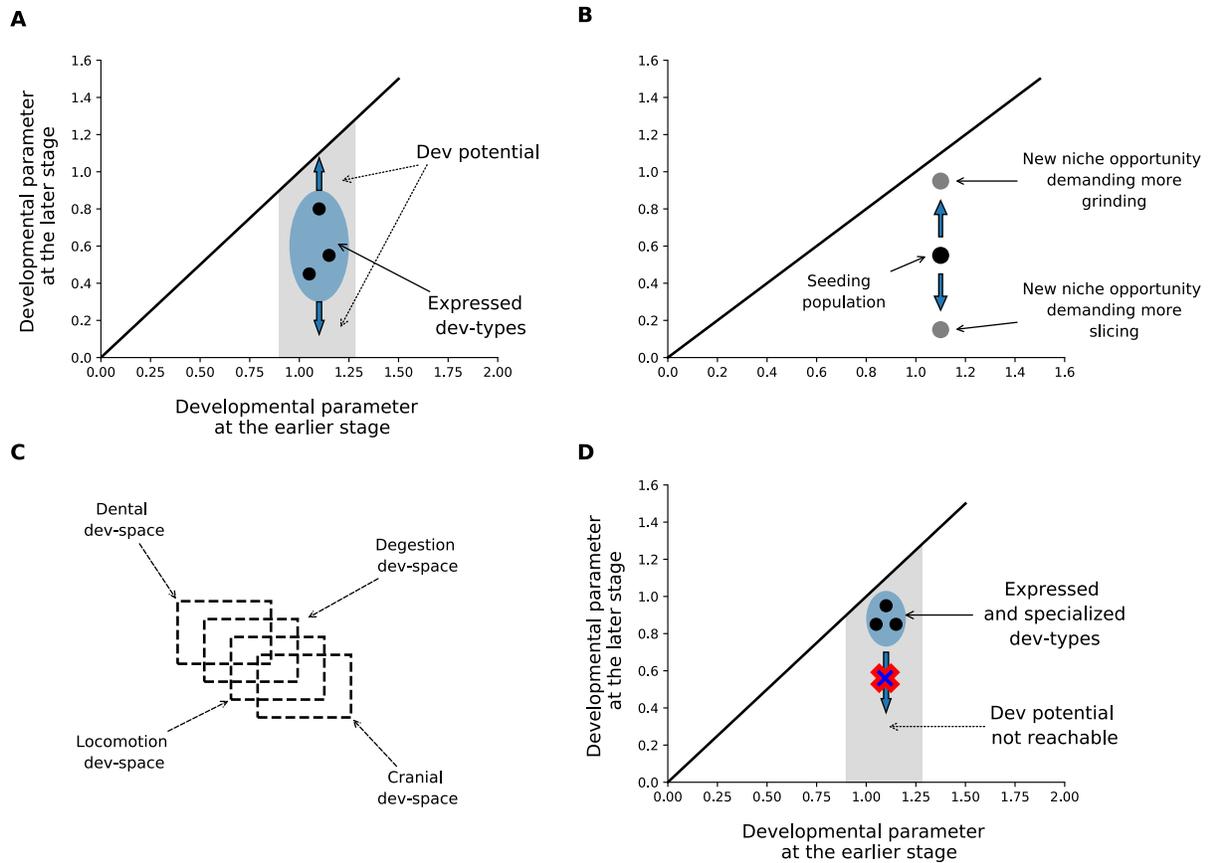

**Fig 5. Schematic plots of the dev-type and dev-space and their application in analysis on adaptive evolution.** (**A**) Developmental potential and expressed developmental variation as suggested by slopes resulting from regressions of adjacent molars. (**B**) Given the dev-type of a seeding taxon or population, the expected evolution of dev-type as new niche opportunities emerging from different directions. (**C**) Schematic representation of the hyperdimensional property of dev-space. (**D**) Schematic plot showing that previous potential dev-types become unreachable as dev-types become specialized.

Above is a simplified analysis of how developmental variables evolve in response to selection. Because many morphological traits are dynamically connected, both functionally and developmentally, dev-spaces from various traits are interconnected with each other, forming a dynamic hyperdimensional graph (Fig. 5C). When a taxon/population is adapted to a specialized niche for long enough, the corresponding morphological traits tend to become more specialized, simultaneously constraining the developmental variability of other traits, and vice versa (*57*). A specific example can be drawn from the specialized folivory adaptation of colobines, which possess a unique digestive system among primates and an exclusive arboreal lifestyle transited from a semiterrestrial habitat with more frugivorous dietary composition (*46, 47*). Although there is no fossil record for the digestive system, it is reasonable to assume that earlier forms (e.g., the Miocene ones) had a more general digestive adaptation (*47*). Nevertheless, because of the specialized digestive system of colobines at present, those potential dev-types locating underneath, though possessed in the past, are unlikely to be regained in the future even if selected for (Fig. 5D).

### *V-4. Generalist vs specialist and evolvability channel*

As previous studies pointed out, by analogy with bet-hedging strategy and the Modern Portfolio Theory, generalists tend to be more resilient to environmental changes compared to specialists (*56, 58-61*). Using analysis of the variability of dev-types, a schematic plot is presented (see fig. S16 and supplementary text) to explain why savannah baboons eventually outperformed geladas in terms of distribution and abundance, although the latter were once as competitive as the former (*47, 62*).

Regarding transitions between different dev-types, or the underlying morphotypes, as discussed above and reported by previous studies (*57, 63*), some are possible while others are less likely or impossible. Here I use a term, evolvability channel, to describe the connectivity between different dev-types. Three major types of evolvability channels can be distinguished: **Type 1**: it is easier for the underlying morphotypes at one island to evolve into that at another, whereas the reverse is more difficult (but possible) to achieve; **Type 2**: it is possible for the underlying morphotypes at one island to evolve into that at another, whereas the reverse is not expected; **Type 3:** it is not expected for the underlying morphotypes at one island to evolve into that at another, nor for the reverse. As irreversibility should be considered as a gradient (*64*), these three types should be viewed as a continuous spectrum instead of discrete states. Using the example of mammalian intermolar inhibitions, evolvability channels are further elaborated (see fig. S17 and supplementary text).

## VI. The combined property of certainty and uncertainty across developmental stages ------- An essential cause for phenotypic variation

### *VI-1. Schematic representation of the combined property*

Given multiple segments, according to the developmental sequence, they are denoted as S1, S2, S3, and S4, such as the case for molars. Within S1, the earlier and later developed parts are denoted as S1: Part A and S1: Part B, respectively (noting that such division can be similarly applied to other segments as well). As shown in Fig. 6, going through developmental stages from earlier to later, segments (or parts) covary, and the covariation may occur across hierarchies. The inhibition intensity of an earlier pair partially determines that of a later pair that overlaps with the former by a segment/part. If inhibition of the former is higher than the threshold value, then inhibition of the latter becomes so strong that the latest developed component will not initiate. Conversely, inhibition of the latter may vary across a wide spectrum.

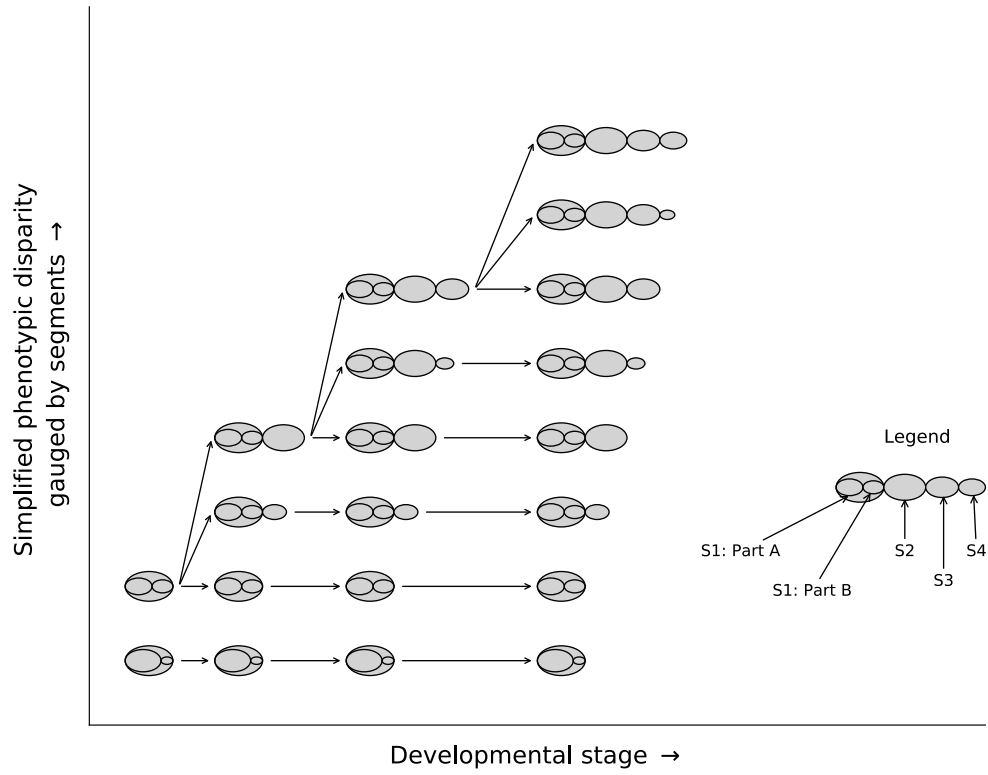

**Fig 6. Schematic presentation of the combined property of certainty and uncertainty across developmental stages.**

*VI-2. Development as a fuzzy translation of the genetic and environmental factors*

Developmental instability indicates that given an individual organism, the development of repeated parts, such as the left and right sets of segments for bilateral organisms, are expected to differ between each other, even though the genetic information is identical and the environmental information may be, though not necessarily, identical (*65, 66*). The combined property of certainty and uncertainty presented here partially explains why developmental instability can be surprisingly high in some cases (e.g., the common occurrence of substantially varied left and right counterparts in bilateral organisms). This property plays an essential role in generating phenotypic variation, as schematically illustrated in Fig. 7 and further elaborated below.

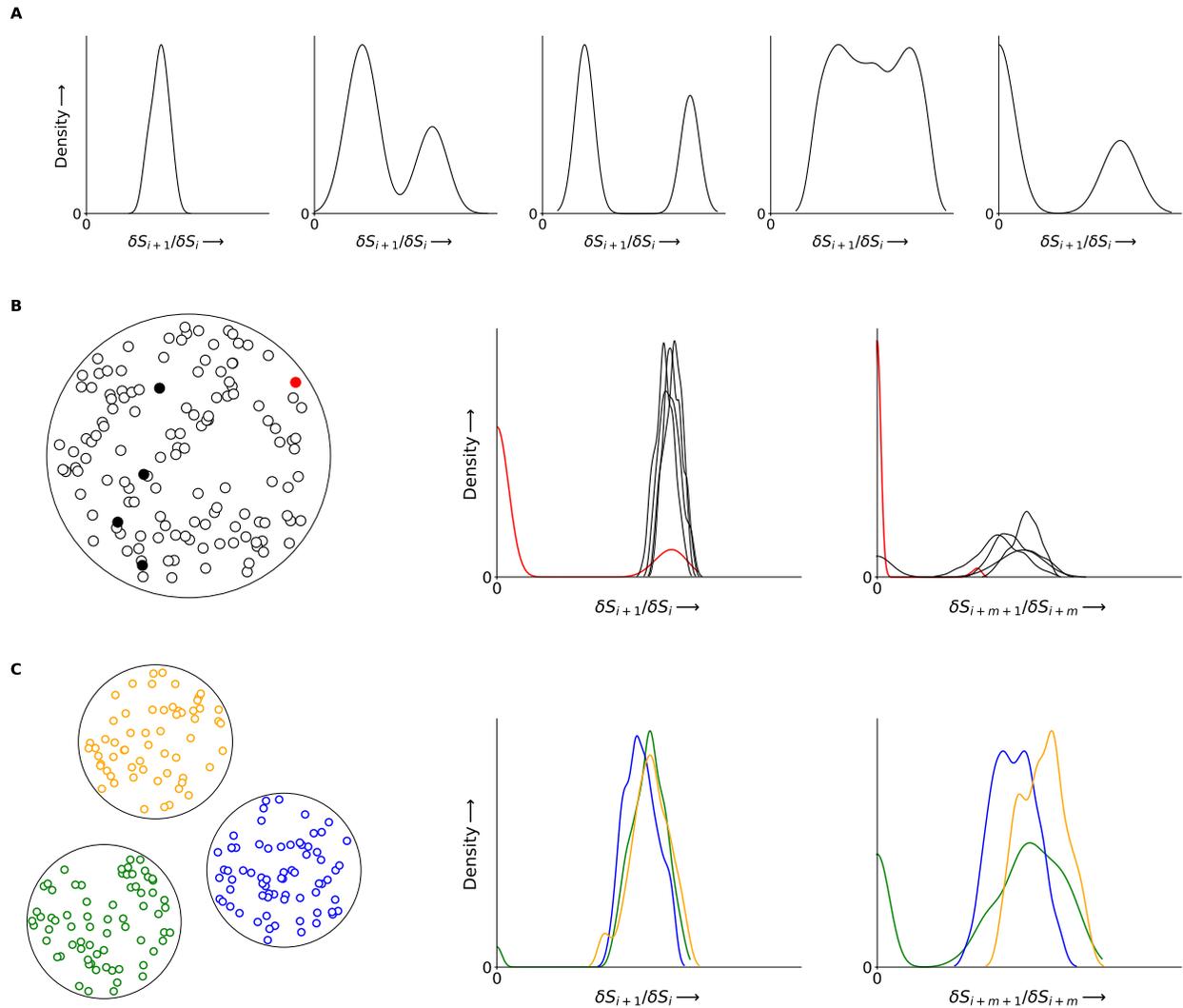

**Fig 7. Schematic illustration showing how the combined property of certainty and uncertainty in development contributes to phenotypic variation.** (**A**) Examples of KDE plots for the inverse of intersegmental inhibition. (**B**) KDE plot (see above) comparison among randomly selected individuals for an earlier and a later stage respectively. (**C**) KDE plot (see above) comparison among populations for an earlier and a later stage respectively.

*VI-2-A. Left vs right:*

Given a set of developmental segments, using $\frac{\delta s_{i+1}}{\delta s_i}$ to denote the inverse of inhibition intensity for a randomly chosen developmental stage (see notations in Section III-3), the KDE plot can be of any form, such as unimodal (symmetrical or skewed), multimodal, gapped, uniform, a fixed value at null, or a combination of a fixed value (i.e., null) and a fluctuating value, etc. (Fig. 7A). Such a free distribution pattern, unpredictable by default, rather than a certain value that can be predicted beforehand, reflects the intrinsic property (i.e., uncertainty) of intersegmental inhibition, which results in a high level of developmental instability seen in natural organisms (*66*).

*VI-2-B. Individual vs individual*:

For individuals within the same population, there are limited genetic differences and presumably limited environmental differences among each other. Even though developmental variances at earlier developmental stages (e.g., anterior intersegmental inhibitions) are expected to be low, those at later stages (e.g., posterior intersegmental inhibitions) may be significantly higher. As for specific individuals, due to the property of uncertainty, some individuals may possess traits that deviate far from the population mode, especially for features developed at later stages, even within populations adapted to narrow ecological niches (Fig. 7B).

*VI-2-C. Population vs population*:

When populations are compared, there are considerable genetic and environmental differences. The congregated KDE of developmental variables (e.g., inhibition intensity) of each population, however, are not expected to differ substantially among each other in general. The differences are expected to correlate with ecological parameters (Fig. 7C). Here too, disparities at later stages tend to be distinctly higher than those at earlier stages, consistent with previous findings such as Butler's Developmental Field Theory (*67*). The patterns shown in the congregated KDE plots also suggest that when only the mean values of developmental variables are examined, interpopulation (or interspecific and so on) disparities are undervalued, like the case of the original ICM versus the generalized ICM, as further elaborated below.

### VI-3. Explanatory power of the generalized ICM versus that of the original ICM

Noting that the generalized ICM is not entirely incompatible with the original ICM, they are applicable to different contexts. Equation (4) presents expected phenotypic variation ranges under development alone, which are subsequently filtered through selection during the evolutionary process (*4*). As a result, the utility curve for the mean values across homogeneous segments tends to follow the original ICM equation when viewed at a higher taxonomic level, a pattern observed across multiple features of diverse organisms (*35, 68, 69*). Nevertheless, because it only captures the broader congregated signal (i.e., the means of higher taxonomic taxa) for phenotypes shaped by selection (Fig. 2), this specific utility curve does not hold when viewed at lower taxonomic scopes or when the segments are heterogeneous (Figs. 2 and 4), as observed in previous studies *(13-20, 53, 55)*.

## VII. Development as a system composed of multi-hierarchical tuning controls

As a simplified illustration (Fig. 8), activators/inhibitors dynamics of earlier developed features can either prevent the initiation of later developed features, leading to a certain outcome, or make the initiation as well as the magnitude of later developed features undetermined, leading to an uncertain outcome.

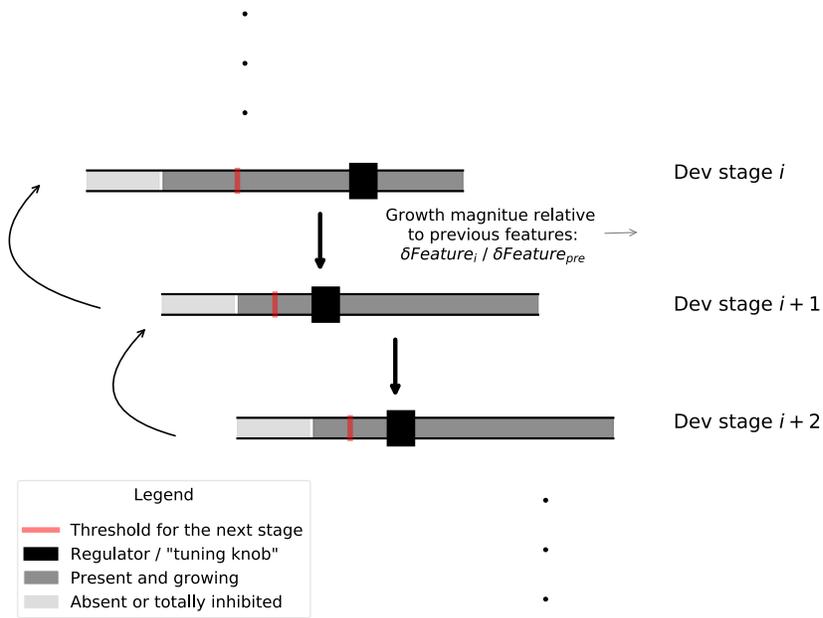

**Fig 8. Schematic illustration for the hierarchical developmental tuning controls across stages.**

This extended form of the hypothesis is useful in interpreting the continuing emergence of novel features that have fostered the broader trend of increasing complexity in evolution (*70, 71*). In the evolution and diversification of mammals, among other features, the rise of new cusps is critical for generating novel phenotypes in correspondence with selection opportunities/pressures. The development of hypocone across multiple taxa during the late Cretaceous (*72, 73*), for instance, is regarded as a key innovation that set the stage for the adaptative radiation of Cenozoic placental mammals (*74*). The addition of hypocone substantially expands the variety of novel molar morphologies that successively arose from the basal four cusps archetype (*75, 76*).

As Turing's Reaction-Diffusion model (*36*) indicates, initiation of later formed cusps are inhibited within a certain range surrounding earlier formed cusps (radii of the inhibition zones are determined by parameters of the model) (*77*). Accordingly, emergence of novel cusps requires sufficient distance from earlier formed ones. The patterning cascade model, derived from studies on seal teeth, suggests a correlation between intercusp distance and cusp number (*78*). Empirical studies on the Carabelli cusp and other accessory cusps of human teeth, however, found that the patterning cascade model cannot accurately explain the presence/absence of later formed cusps when tested on teeth more complicated than seal's (i.e., beyond one dimension) (*79, 80*). In other words, sufficient space may or may not lead to the emergence of new cusps. The same relationship has also been found in other cases. For example, the development of hypocone requires a decent size of the posterolingual cingulum, but the reverse does not always hold (*81*).

As suggested by the hierarchical tuning mechanism (Fig. 8), achieving complexity is extremely difficult from chance alone. For example, it is required to simultaneously adjust multiple developmental pathways to generate mutant variants with complicated tooth forms, still differing from the morphologies found in nature (*82*). Albeit infinite in all cases, the total quantity of potential phenotypes grows substantially as developmental stage accretes. Hence, if the targeted form is complicated, then it is unlikely to retrieve a phenotype in lab that can also be found in nature. Conversely, if the targeted form is simple, then retrieving the natural phenotypes becomes much easier.

Meanwhile, although generating complicated phenotypes is difficult from the developmental perspective alone, selection will significantly boost the chance in evolution when such forms become adaptive (*76*). As a consequence, although getting simplified phenotypes is developmentally easier (*83*), increasing complexity is the trend in evolution. Finally, given that change of environment is unpredictable and thus the direction of selection is fluctuating, the combined property of certainty and uncertainty across developmental stages, together with the property of randomness from genes (*84-86*), plays an essential role for organisms to minimize the chance of going extinct by maintaining the capability of generating a diversified spectrum of phenotypic stocks.

VIII.  Conclusion

As stressed in the beginning, Darwin's statement regarding our ignorance of the "*laws of variation*" remains valid as of today. Along with prior studies contributed by other researchers, the hypothesis and related discussions presented here, hopefully, might unveil the "*laws of variation*" by a little bit. As a noncomplete list, some key points are emphasized below:

1. The generalized ICM, as tested using both development data and evolutionary data, may be applicable to a broad range of organisms (Sections I - III).

2. The original ICM formula, corresponding to a specific utility curve, results from selection. It is only applicable to homogeneous segments when viewed at a high scale. If either of the conditions is not met, then proximal distal trade-offs become imbalanced, leading to a shift of the utility curve (Sections II and IV).

3. The combined property of certainty and uncertainty across developmental stages plays a key role in generating phenotypic variation (Section VI). This property is essential for organisms to survive through fluctuating ecological settings during the evolutionary process (Section V).

4. At a higher level, development can be viewed as a multi-hierarchical system composed of multiple tuning controls chained together across earlier to later stages. The tuning range, rather than any specific variable except for null, of a given stage is determined by the tuning variable adjusted at the immediately preceding stage (Sections III, VI, and VII).

**Methods**

Linear regressions in the present study are conducted using the ordinary least-squares (OLS) method, which is different from the reduced major axis (RMA) used in some past relevant studies (*10, 20*). As Smith (*87*) has pointed out, in biological studies, a criterion on choosing between OLS and RMA is to see whether the variables are symmetrical. If they are, then it would be justified to use RMA; if not, then OLS should be preferred over RMA. In the present study, the earlier developed components are always treated as independent variables whereas the later developed components are always treated as dependent variables. Importantly, there is indeed a biological causal relationship between the independent and dependent variables. Hence, OLS rather than RMA is the correct choice here for conducting linear regressions. Full materials and methods are presented in the Supplementary Materials.

89. V. J. Lynch, Is there a loophole in Dollo's law? A DevoEvo perspective on irreversibility (of felid dentition). *Journal of Experimental Zoology Part B: Molecular and Developmental Evolution*, (2022).

**Acknowledgments:** First and foremost, I thank Professor Richard Carr Fox (1933-2022). Professor Fox maintained the true spirit of a scientist. In my own opinion, the culture of doing science (in the field I am in, whatever that is) has become more and more like conducting political campaigns. Many scientists, whether willingly or not, have prioritized their goal of seeking congruence with the majority, or more precisely, those influential players. In contrast, throughout his whole scientific career (a.k.a. life), Professor Fox never deviated from the path of seeking truth regardless of the critiques, mostly (if not always) without solid support, from those prestigious peers. I have seen the true dignity of a scientist in Professor Fox, and I feel extremely fortunate to have become acquainted with him and to have received his mentoring and help on numerous occasions. Next, for the specific study presented here, I am grateful to Jukka Jernvall for generously granting me the access to the original data of the 2007 *Nature* paper on the Inhibitory Cascade Model *(10)* (table S1), and I thank Debbie Guatelli-Steinberg, John Hunter, Greg Wilson, and Christoph Zollikofer for discussions and help. I am indebted to my former colleagues as well as former/present/future friends, Wataru Morita, Ryan Murray, Ali Vahdati, and Jody Weissmann for discussions, supports, and inspirations. I am also grateful for past studies contributed by numerous researchers including but not limited to Peter Lucas, Michael Plavcan, Kathryn Kavanagh, Blaire Van Valkenburgh, and Masakazu Asahara, without whose prior contributions the study here would not have been possible.

**Funding:** The primary funding source is the author's personal fund. In addition, this study is partially funded by the Swiss National Science Foundation Grant #CR32I3_166053 (awarded to Christoph Zollikofer).

**Author contributions:** YZ contributed to all parts from conceptualization to writing.

**Competing interests:** The author declares that there are no competing interests.

**Data and materials availability:** All data are available in the main text or the supplementary materials.
**Supplementary Materials**

Materials and Methods

Supplementary Text

Figs. S1 to S17

Tables S1 to S3

References (*88*–*89*)

# Supplementary Materials for

## Combined certainty and uncertainty across development frees phenotypic variation in evolution

Yue Zhang

Correspondence to: renaissance.of.truth.chasing@gmail.com

**The PDF file includes:**

    Materials and Methods
    Supplementary Text
    Figs. S1 to S17
    Captions for Tables S1 to S3
    References *(88-89)*

**Other Supplementary Materials for this manuscript include the following:**

    Tables S1 to S3 (zipped .xlsx file)

**Materials and Methods**

Materials on developmental data
The data are from the mutant samples of the original experiment associated with the original ICM paper *(10)*. Please refer to the original publication for details of the experiment. The data are presented in table S1.

Materials on anthropoid data
As stated in the main text, the data are from Plavcan's dissertation *(22)*, also previously published in the supplementary table by Carter and Worthington *(20)*. According to Plavcan, regarding intraspecific samples, some were collected from one single site whereas some were collected from multiple sites, and data from multiple subspecies were included in some cases *(22)*. Here, for the purpose of minimizing sampling bias, only species with a moderate samples size (n > 30) were included (table S3).

Materials on carnivoran data
These data are from the study presented by Asahara and colleagues *(14)*. Please refer to the original publication for details.

Materials on the hindlimb data
These data are from the dataset compiled by Young *(28)* from past studies (see details and references therein).

Notation on mammalian molars
In the texts, mammalian molars are denoted as M1, M2, M3, etc.. The underlying data, including developmental and evolutionary, are typically from lower molars. Noting that lower molars were not denoted as m1, m2, m3 here, contrary to the conventional notation in some fields such as paleontological studies, the reason is to maintain consistency with other ICM related papers (e.g., notations used in the original ICM paper *(10)*).

**Supplementary Text**

Utility line versus the original ICM equation
As mentioned in the main text, the utility curve $U(\alpha) = 2 - 1/\alpha$, in which $\alpha$ denotes $\delta M3/\delta M2$, and $U(\alpha)$ denotes $\delta M2/\delta M1$, corresponds to the ICM equation $M2 = \frac{1}{2}(M1 + M3)$. Below, some additional explanations are provided which might be helpful for some readers. Readers with quantitative background could choose to skip this subsection to avoid redundancy.

From the equation of the utility curve, we can get $\frac{\delta M3}{\delta M2} = 2 - \frac{\delta M1}{\delta M2}$, which is equivalent to $\delta M3 = 2\delta M2 - \delta M1$, so we have $\delta M2 = \frac{1}{2}(\delta M1 + \delta M3)$, and vice versa. Noting that the utility curve is preferably to be derived from slope data, as in the case of anthropoid molar data analyzed in the main text. Hence, the utility curve $U(\alpha) = 2 - 1/\alpha$ corresponds to the original ICM equation, but these two equations are not equivalent.

Size ratios versus slopes
Previous studies mostly used size ratios instead of slopes. As explained in Section II-1, slopes of evolutionary data are informative for inferring intermolar inhibitions during development. In contrast, size ratios as final products of development, cannot be used as direct references for intermolar inhibitions. Rather than the cases seen in lab experiments, in natural samples, variations in factors such as the initiation time, developmental duration, and calcification time all affect the final molar size ratios *(17, 20)*. Hence, slopes should be used to infer intermolar inhibitions. Nevertheless, slopes are not always retrievable, especially when the sample size is not sufficient. For the case of anthropoid molars as analyzed here, although showing considerable deviation compared to the scatterplot from slopes, the mean size ratios of included species are all located beneath the line $y = x$, agreeing with the broader pattern revealed by slope data (fig. S12 and Fig. 2). Accordingly, when slope data are not available, in order to estimate the intermolar inhibition intensities in development, mean size ratios of natural samples might be tentatively used as a secondary option.

Connection between the generalized ICM and non-concave outline for homogeneous segments
For simplification purpose, three adjacent segments, denoted as $S_1$, $S_2$, and $S_3$, respectively, are represented by circles, and the radii of these three circles are denoted as $r_1$, $r_2$, and $r_3$, respectively (fig S14). The schematic drawing shows a threshold condition in which $r_1 < r_2 < r_3$ and the outline connecting $S_1$, $S_2$, and $S_3$ approximates a straight line. After drawing some auxiliary lines as shown in the figure, it is seen that:
$$\frac{|CE|}{|BE|} = \frac{|BF|}{|AF|}$$
Hence, $\frac{r_3-r_2}{r_3+r_2} = \frac{r_2-r_1}{r_2+r_1}$. After some elementary transformations (omitted), this leads to $r_3 r_1 = r_2^2$, which equals to $\frac{r_3}{r_2} = \frac{r_2}{r_1}$. From this threshold condition, if the relative size of $S_3$ increases, then $\frac{r_3}{r_2} > \frac{r_2}{r_1}$, corresponding to a concaved outline. Conversely, if the relative size of $S_3$ increases, then $\frac{r_3}{r_2} < \frac{r_2}{r_1}$, corresponding to a convex outline. Although the example here is discussed using the assumption of $r_1 < r_2 < r_3$, other situations can be discussed in a similar way and are omitted accordingly. Interested readers are suggested to perform the tests themselves.

Generalists vs Specialists analyzed by dev-types
Past studies, especially from ecological perspective, have discussed the pitfall of going specialization by analogy with bet hedging *(58-60)* or the Modern Portfolio Theory in financial research *(56)*. In short, given a population, because each generation's richness depends on that of previous generations, geometric mean rather than arithmetic mean is the key factor. Hence, maintaining a stable quantity, instead of going through high volatility (i.e., wide swing of up and down), is essential to a population's survival in the long term. Accordingly, maintaining the disparity of genetic information, habitats, and geographic ranges outperforms going specialized in the long run *(56)*.

The property of dev-space indicates that a relaxed developmental parameter for the earlier stage (i.e., earlier-later inhibition is not strong enough to make the initiation impossible for the later developed feature) makes the parameter more plastic for the later stage. Therefore, potential dev-types regarding later developmental stages can maintain a relatively broad range in most cases.

However, once the phenotypic trait or its associated traits becomes specialized, potential dev-types may become unreachable, as discussed in the main text. In contrast, generalists still have access to those potential dev-types.

Given two populations, one is specialist and the other is generalist, as plotted in fig. S16. These two populations share the earlier developmental parameter (e.g., size slope for M2 on M1), but differ in the later developmental parameter (e.g., size slope for M3 on M2). One specific example would be geladas as the specialist versus savannah baboons as the generalist, in which case, the specialist feeds almost exclusively on grass, whereas the generalist has a much wider dietary spectrum *(47)*. When niche opportunities of grass feeding increased, such as the case during most of Pleistocene, the specialist (i.e., geladas) increased in abundance and habitat span, but remained specialized of the highly adaptive and restricted dev-types. Fossil record shows that geladas in Pleistocene were more broadly distributed and greater in body size, but also more specialized *(47, 62)*. Because environmental fluctuation is unavoidable, in the long run, specialists are set to decrease its richness or even go extinct (fig. S16). As for geladas, after a limited period of flourishing, they have become endangered. In contrast, their generalist counterpart, Savanah baboons, are more resilient to environmental fluctuation due to the higher variability of dev-types (fig. S16).

Evolvability channel discussed using a specific example of mammalian molars
Because of developmental bias, some morphotypes are expected to be more common than some others, and some morphotypes may not be obtainable at all. Therefore, from the evolutionary perspective, connections among different morphotypes are discontinuous *(7)*. For simplification purpose, the schematic plot here contains two conceptual dimensions, with each representing a conceptual morphological component. As shown in fig S17, continuous morphotypes can be viewed as an island *(7)*. Here, five such islands are schematically illustrated, with arrows used to denote the evolvability channels among these islands.

The specific example used here is about various types of lower molar sizes for placental mammals. If only M1 is present, then the morphotypes are denoted as Island A. If M2 is present but M2/M1 is below the threshold, so that M3 is absent, then the morphotypes are denoted as Island B. If M2/M1 is above the threshold but M3 is absent, then the morphotypes are denoted as Island C. If M3 is present, but smaller or subequal to M2, then the morphotypes are denoted as Island D. If M3 is significantly larger than M2, then the morphotypes are denoted as Island E. In this example, following the sequence A-B-C-D-E, relative magnitude of posterior molars gets enlarged. Noting that this is an oversimplified schematic representation, morphotype islands may be discontinuous across various scales. In other words, the texture of each island is not continuous, with the islands themselves being composed of multiple smaller islands, but only the major islands are discussed below for simplification purpose.

As illustrated here, it is easier for morphotype island D to evolve into other islands, but not the other way around. Morphotype island C is a special case. When M2/M1 is above the threshold value, it is expected to be a two-way channel between the morphotype with M3 present and that with M3 absent. For instance, it has been claimed that *Callimico*, the only member of Callitrichini that possesses M3, experienced an evolutionary reversal in regaining M3 from an ancestral state that had lost M3 *(88)*. Meanwhile, the utility curve indicates that the threshold value for M2/M1 is 0.5 (Fig. 2). The M2/M1 ratio of *Callimico* is above this value (personal observation). Hence, developmentally speaking, such evolutionary reversal can happen because the morphotype island of the subject taxon and that of the ancestor taxon are bidirectionally connected. This is consistent with the hypothesis that irreversibility (Dollo's Law) is expected to hold at some scenarios but not at some others, and thus should be viewed as a gradient *(64)*.

From morphotype island D to that of A and B, examples can be drawn from the repeated evolution of hypercarnivores, presumably from relatively more general morphotypes, whereas specialized hypercarnivores periodically went extinct without the ability to evolve back into more general forms due to the macroevolutionary rachet (i.e., multiple structures get specialized simultaneously so that no structure can evolve back into the more general form without inducing major changes of other specialized structures) *(57, 63)*. For the relationship between D and E, it is a one-way channel similar to the case discussed above. Although it is plausible for a general form in island D to evolve into a specialized form in Island E, of which M3 is substantially larger than M2, the reverse is likely to be restrained (see discussion in the subsection above). For the relationship between C and E, the direct path is closed for both directions (i.e., Type 3), but the evolutionary route is theoretically open through Island D.

Accessibility between morphotype islands A and B fits into the intermediate state between Type 1 and Type 2 as defined above. It is easier to evolve from B to A, whereas the reverse is rare though not impossible, such as the case for Eurasian lynx (*Lynx lynx*), of which a decent proportion of intraspecific variants may possess a tiny though non-typical M2 *(38, 89)*.

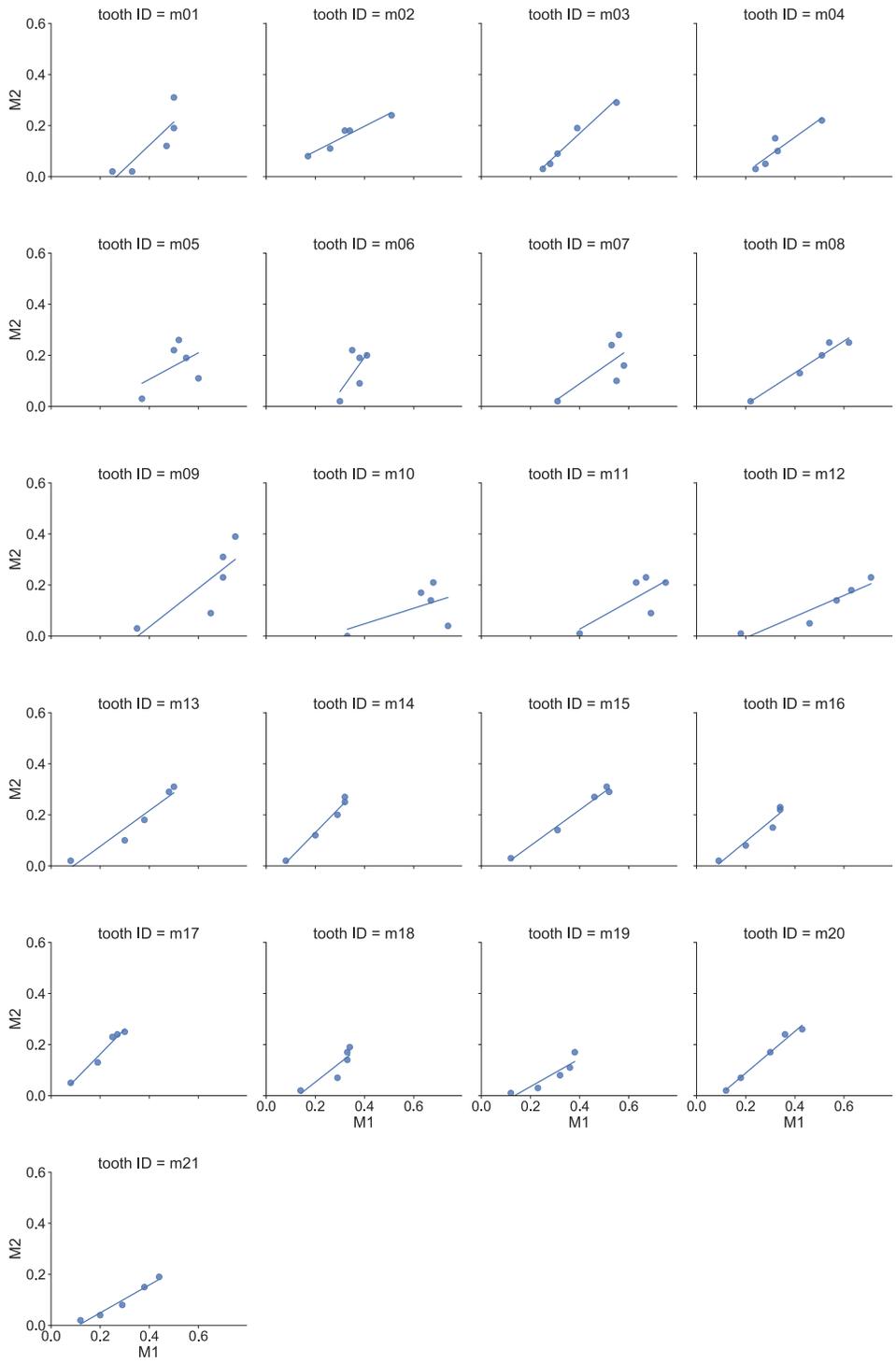

**Fig. S1.**
**Developmental series of M2 vs M1 of each individual tooth *in vitro*.** Refer to the original ICM paper *(10)* for experimental details.

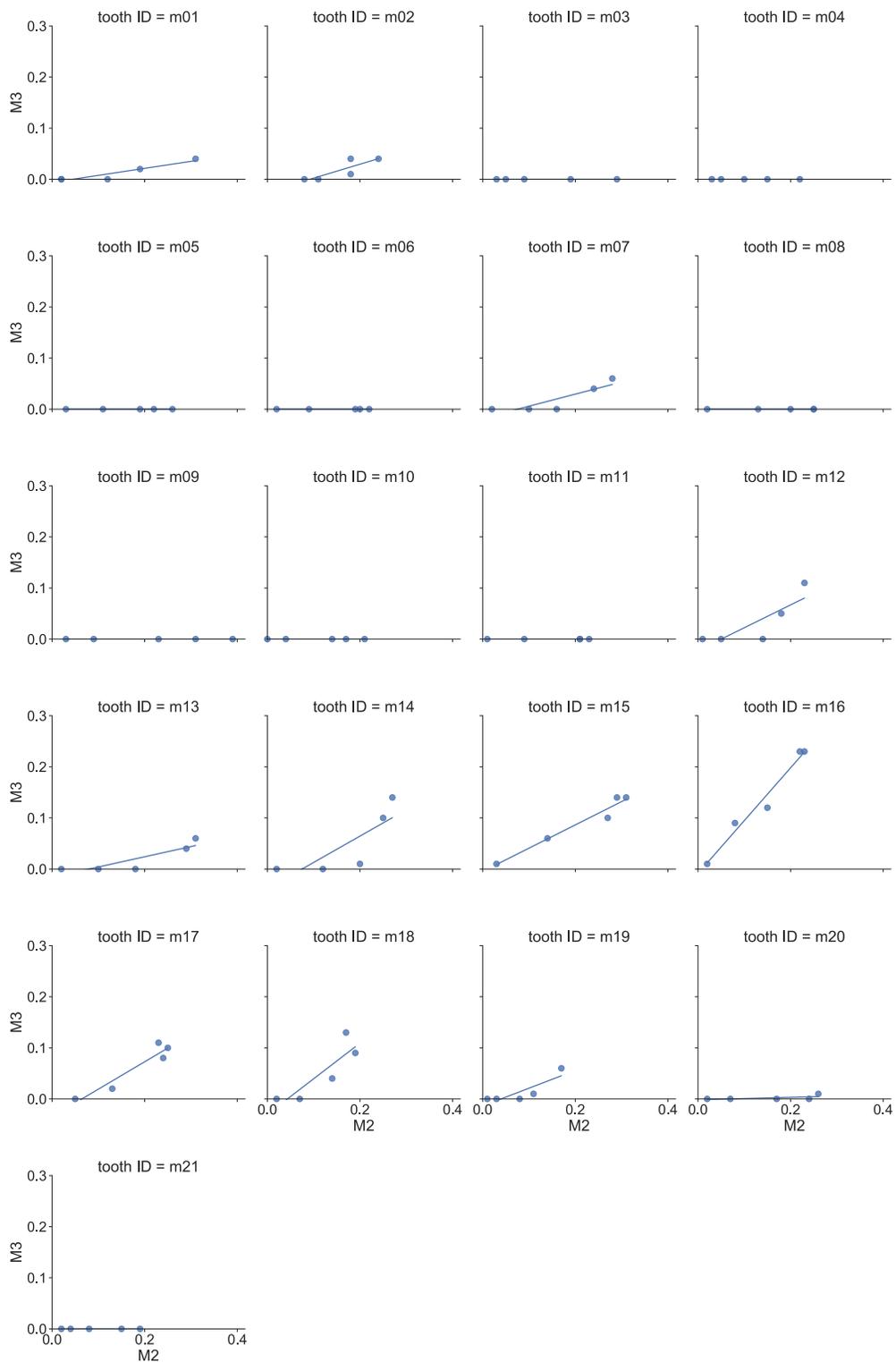

**Fig. S2.**
**Developmental series of M3 vs M2 of each individual tooth *in vitro***. Refer to the original ICM paper *(10)* for experimental details.

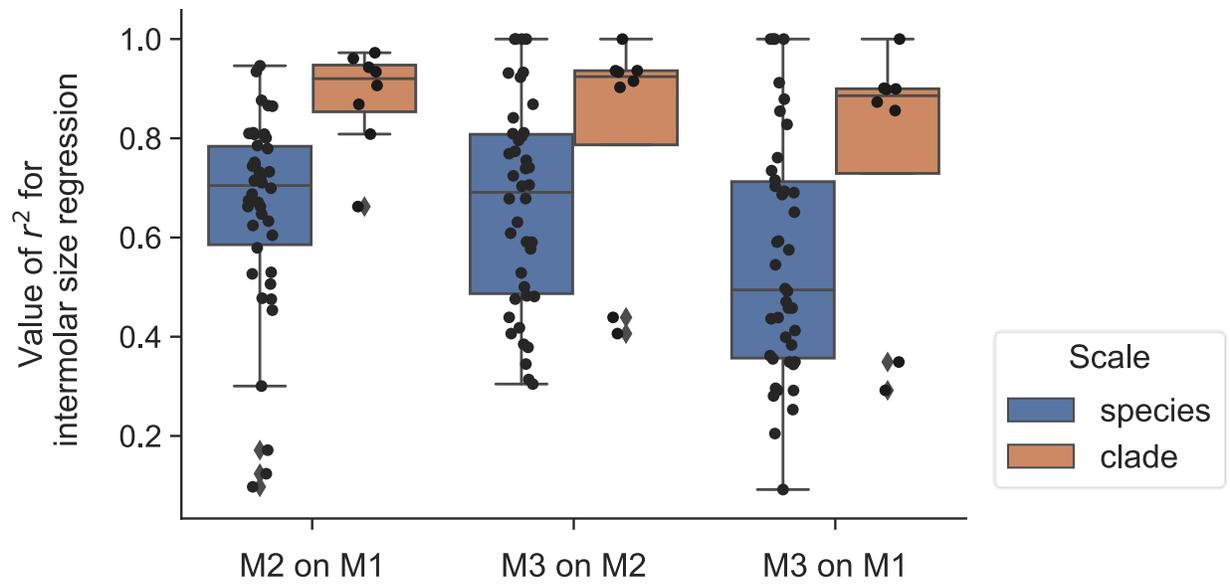

**Fig. S3.**
**Linnear fit of size regressions between molars compared between different taxonomic level.**
Data are from Plavcan's dissertation *(22)*, as reported in Carter and Worthington *(20)*, presented in Table S3.

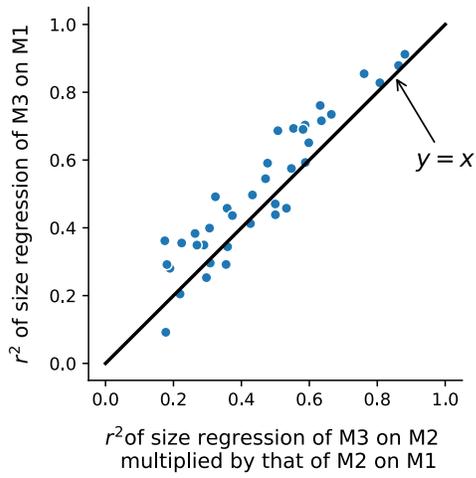
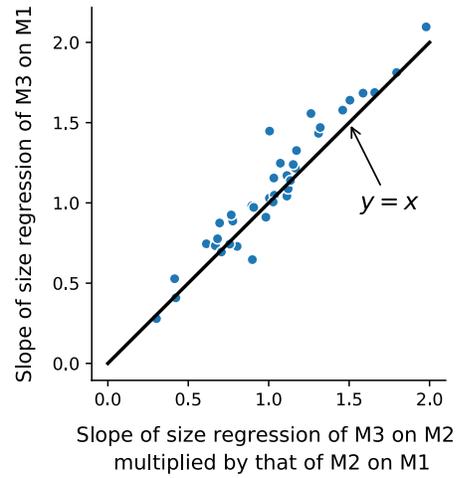

**Fig. S4.**
**Linear fit of M3 vs M1 determined by that of M3 vs M2 and M2 vs M1, as revealed by anthropoid data** *(20, 22)* **(table S3)**.

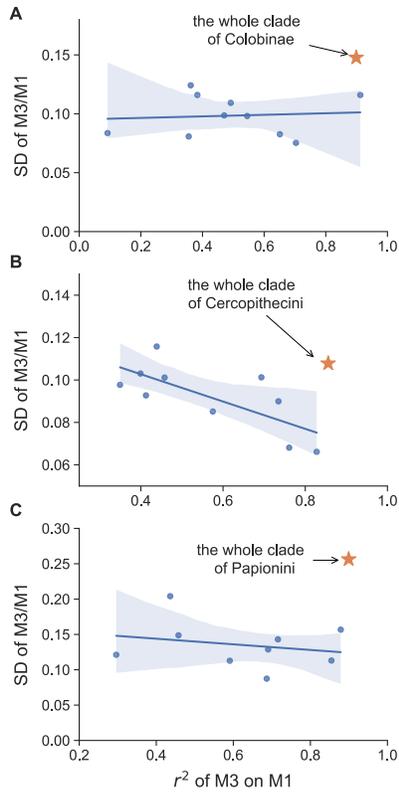

**Fig. S5.**
**Regressions showing the correlation between the linear fit of M3 on M1 and the standard deviation of M3/M1**. Data are from Plavcan's dissertation *(22)*, as reported in Carter and Worthington *(20)*, presented in table S3. (**A**) The regression for species of Colobinae and the whole clade of Colobinae. Shaded area indicates confidence intervals. (**B**) The regression for species of Cercopithecini and the whole clade of Cercopithecini. (C) The regression for species of Papionini and the whole clade of Papionini.

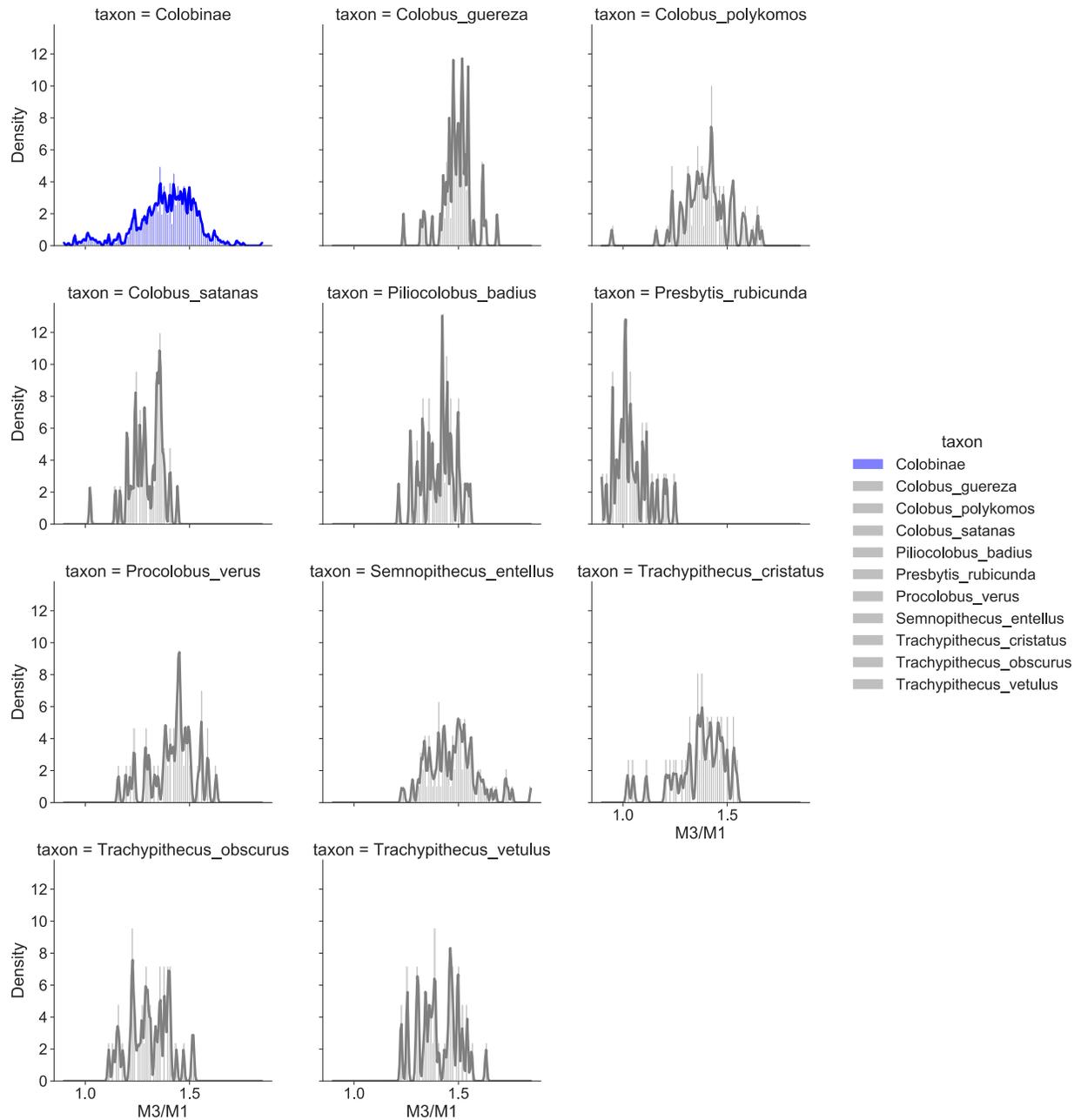

**Fig. S6.**

**Histogram (with KDE) of the M3/M1 ratio among species of Colobinae compared with that of the whole clade**. Data are from Plavcan's dissertation *(22)*, as reported in Carter and Worthington *(20)*, presented in table S3.

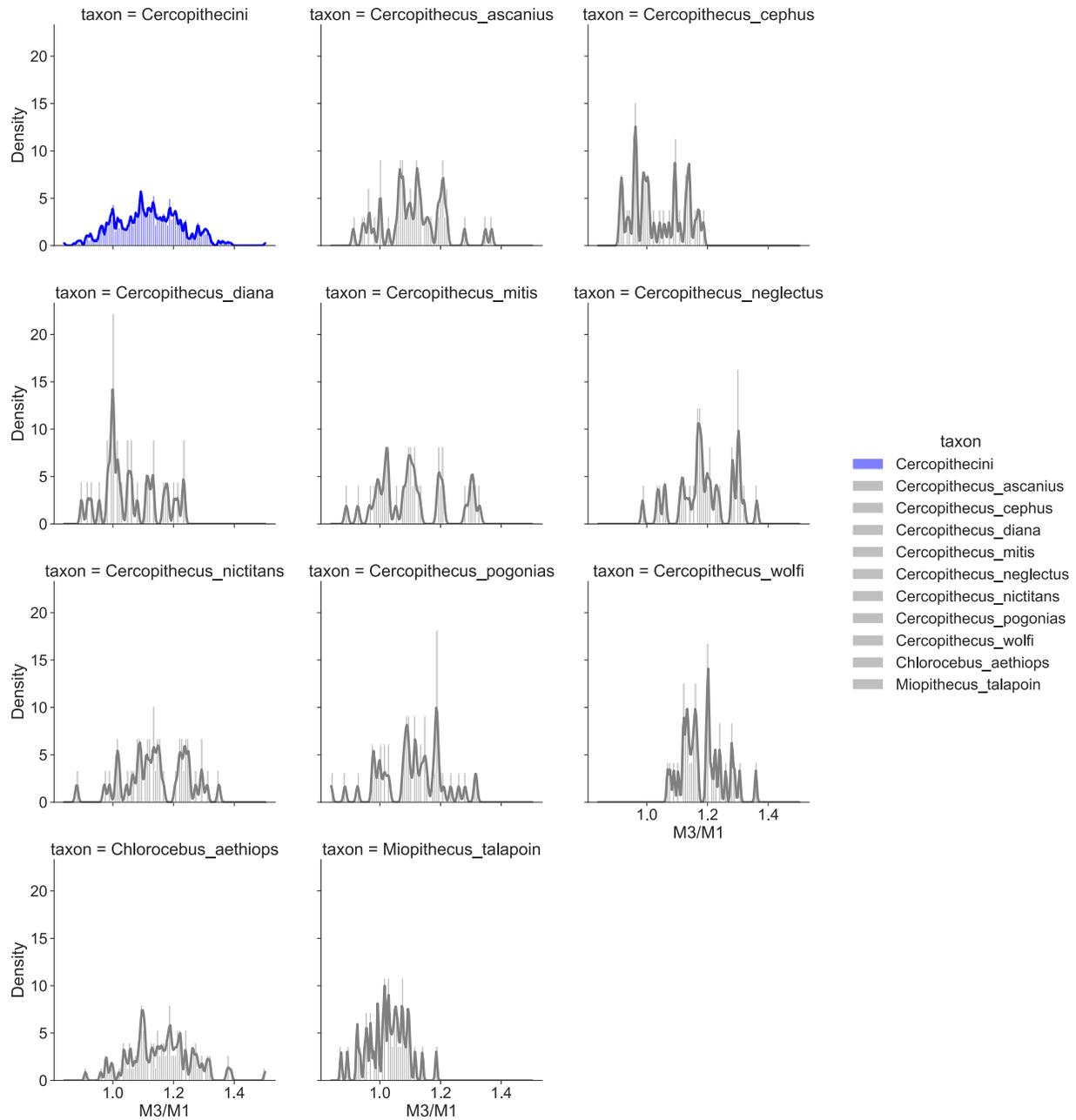

**Fig. S7.**
**Histogram (with KDE) of the M3/M1 ratio among species of Cercopithecini compared with that of the whole clade**. Data are from Plavcan's dissertation *(22)*, as reported in Carter and Worthington *(20)*, presented in table S3.

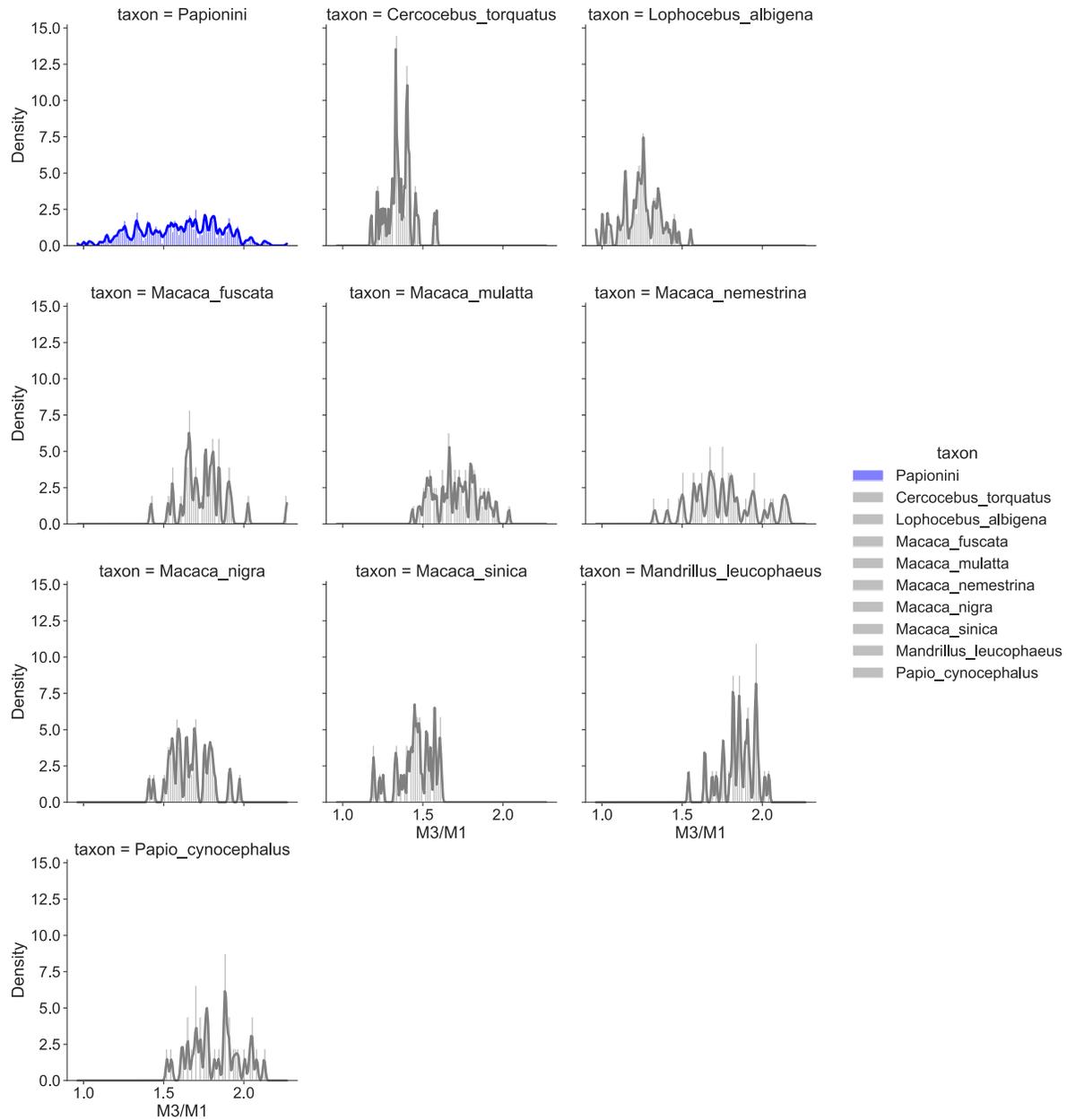

**Fig. S8.**

**Histogram (with KDE) of the M3/M1 ratio among species of Papionini compared with that of the whole clade**. Data are from Plavcan's dissertation *(22)*, as reported in Carter and Worthington *(20)*, presented in table S3.

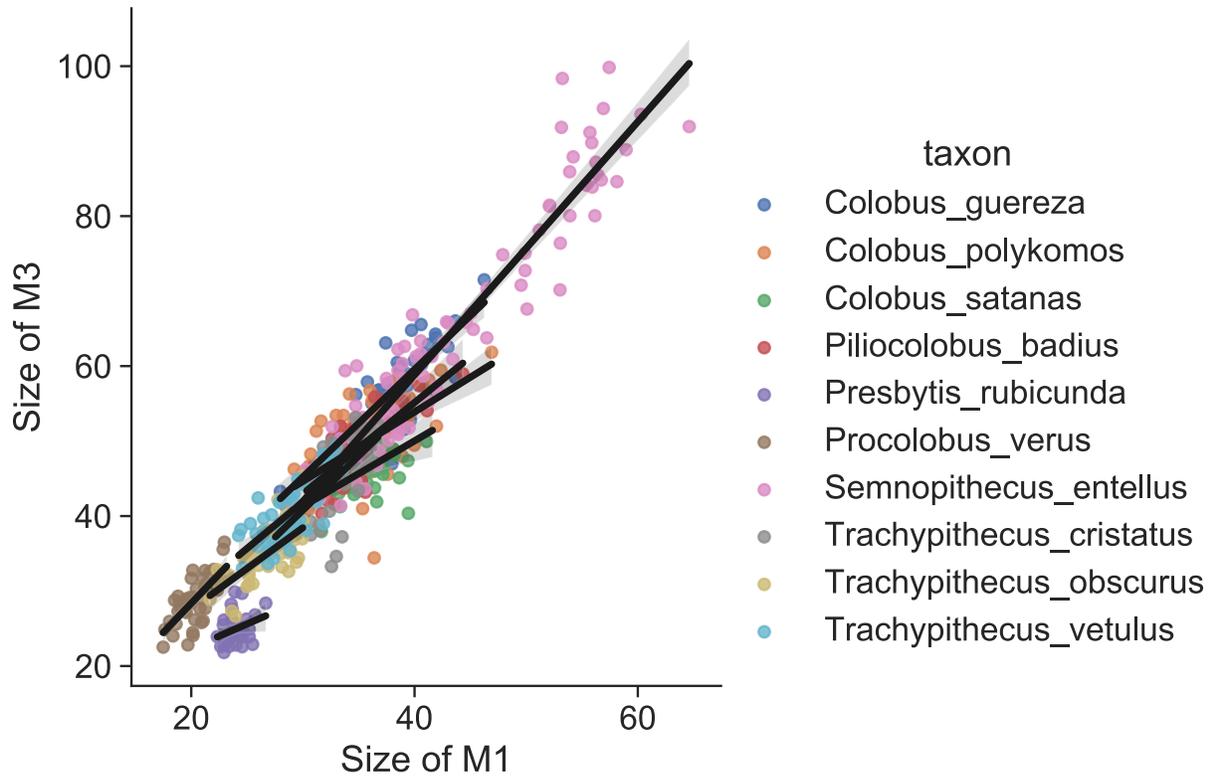

**Fig. S9.**
**Regressions compared between species-level and the clade-level for Colobinae**. Data are from Plavcan's dissertation *(22)*, as reported in Carter and Worthington *(20)*, presented in table S3.

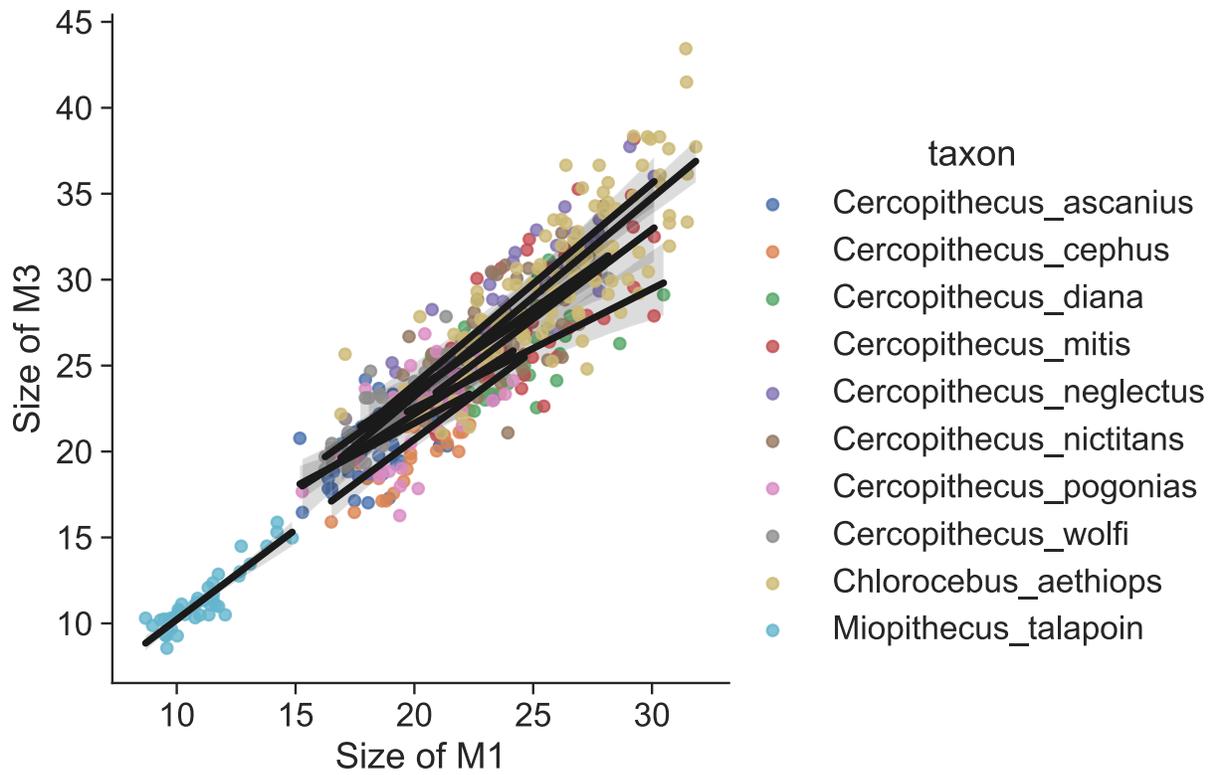

**Fig. S10.**
**Regressions compared between species-level and the clade-level for Cercopithecini**. Data are from Plavcan's dissertation *(22)*, as reported in Carter and Worthington *(20)*, presented in table S3.

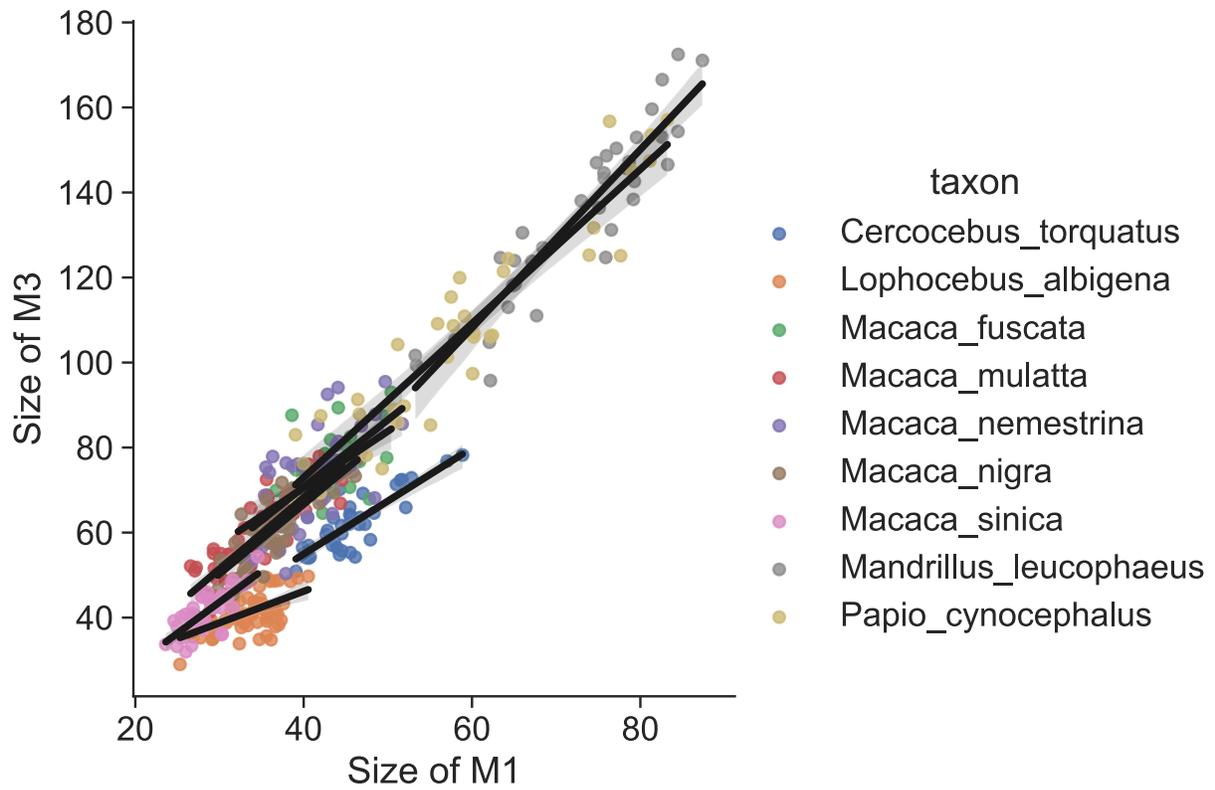

**Fig. S11.**
**Regressions compared between species-level and the clade-level for Papionini.** Data are from Plavcan's dissertation *(22)*, as reported in Carter and Worthington *(20)*, presented in table S3.

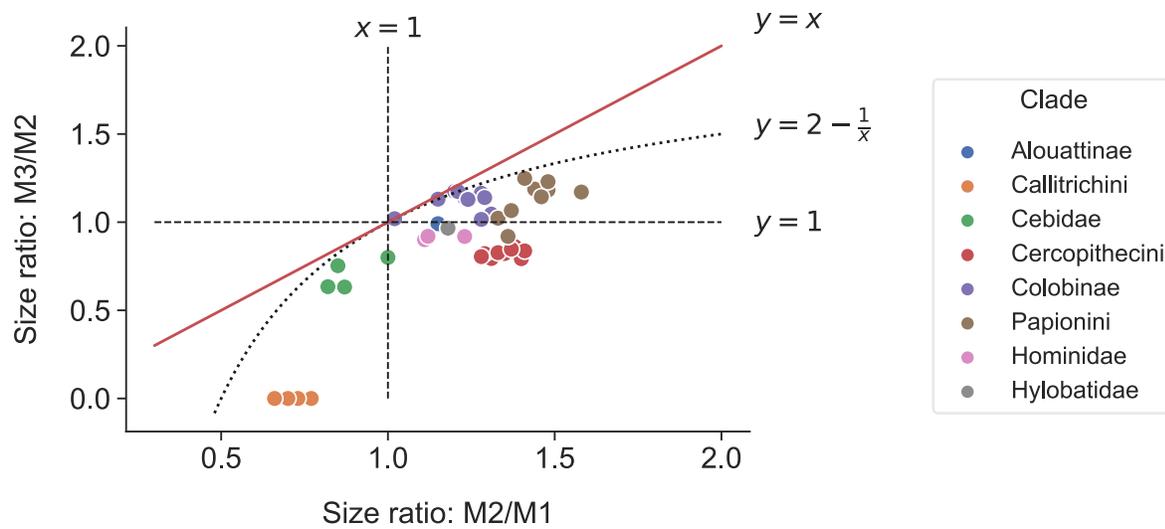

**Fig. S12. Covariation between intermolar ratios of mean sizes for anthropoids.** Species are separated by clade at tribe, subfamily, or family levels. The data are from Plavcan *(22)*, as reported by Carter and Worthington *(20)*.

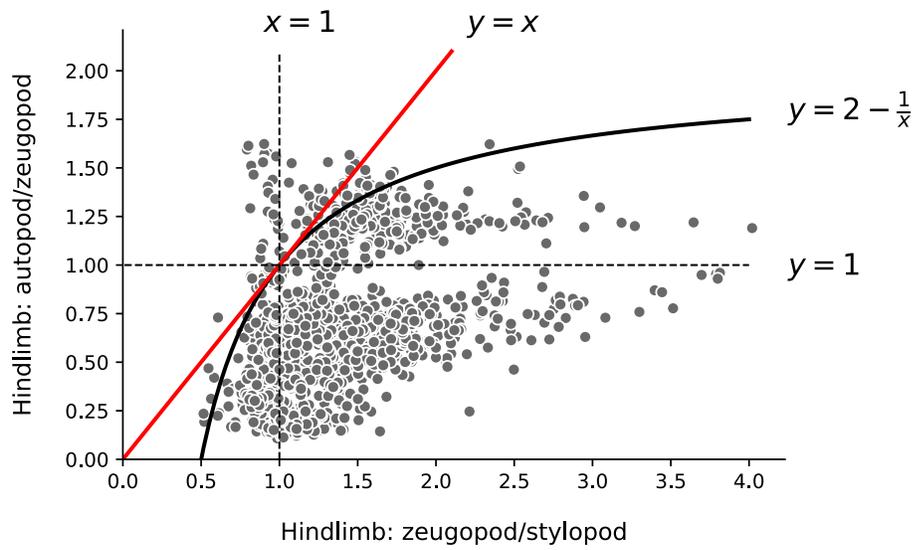

**Fig. S13. Plot of segmental size ratios of hindlimbs ($N_{limbs}$=1452, $N_{taxa}$=1354).** The data are species-level population-averaged measurements compiled by Young *(28)*.

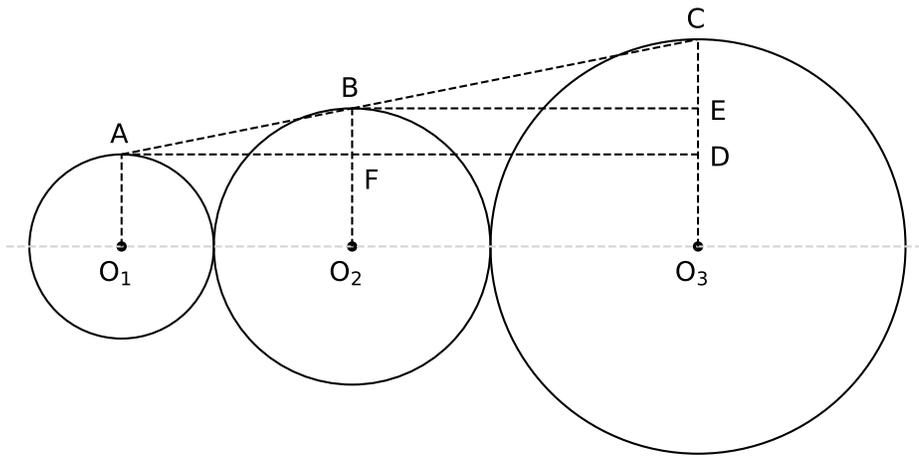

**Fig. S14. Schematic illustration to explain outline curves in relation to the generalized ICM.** See text for details.

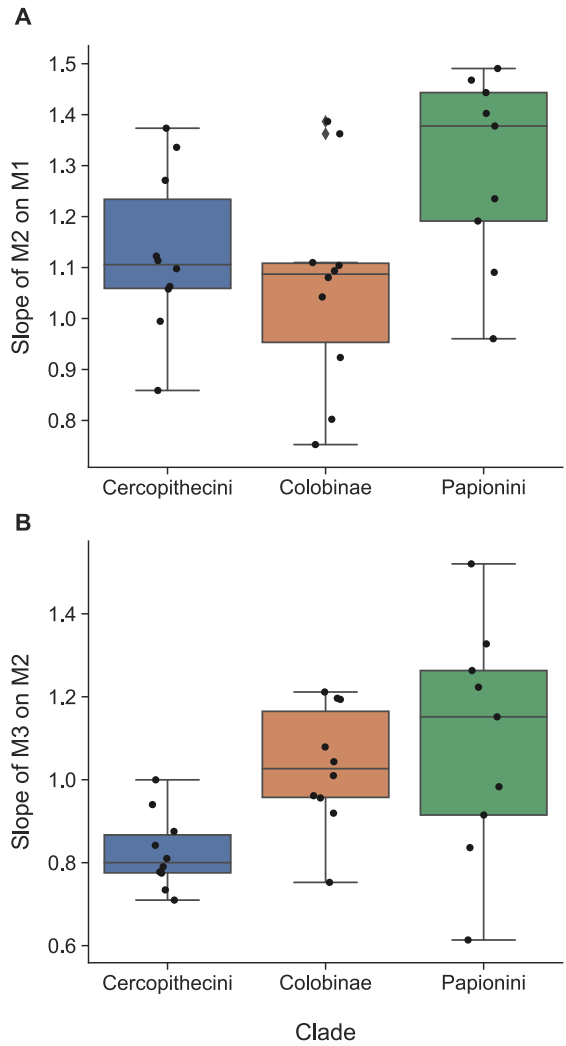

**Fig. S15.**
**Comparison of the slopes resulting from regressions of adjacent molars among the three clades of Old World monkeys**. Data are from Plavcan's dissertation *(22)*, as reported in Carter and Worthington *(20)*, presented in table S3.

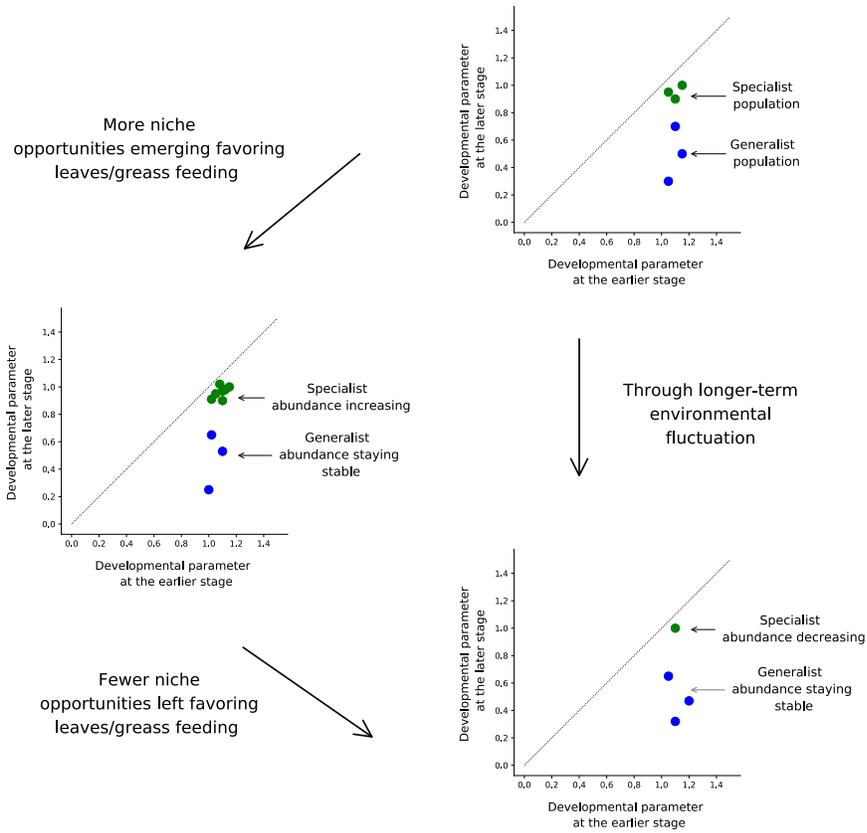

**Fig. S16.**
**Specialists vs generalists facing short-term and long-term environment fluctuations, as viewed from developmental perspective**.

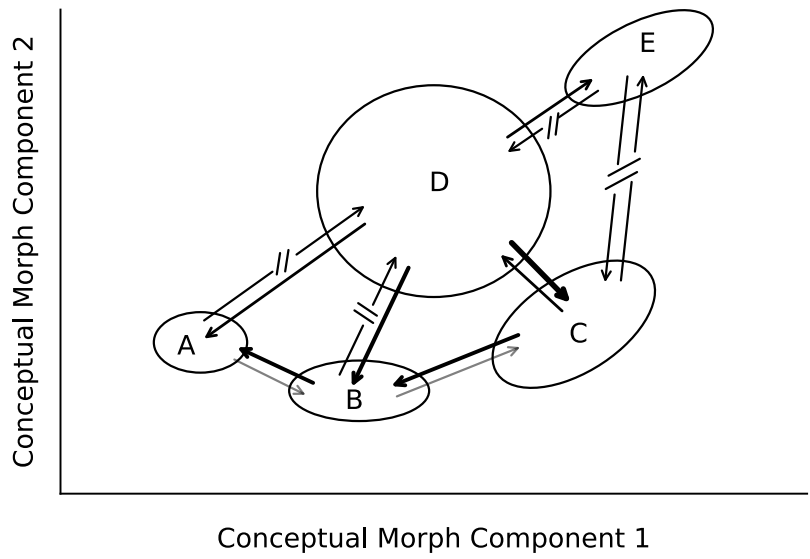

**Fig. S17.**
**Schematic representation of the evolvability channels between disjoint morphotype islands.**
The thickness and darkness of the lines indicates the easiness of evolving from one morphotype island to another. See text for discussion in detail.

**Supplemental Tables**

Supplemental tables are all located in a separate zipped .xlsx file.

**Table S1.**
Development data from the experiment of the original ICM paper *(10)*.

**Table S2.**
Slopes (and $r^2$ values) of the developmental data.

**Table S3.**
Anthropoid data including the slopes and $r^2$ values.